\def\etal{{\it et al. }}
\newcommand{\angstrom}{\AA}
 \documentclass[final,3p,times]{elsarticle}
\usepackage{amssymb}
\usepackage{amsmath}
\usepackage{float}
\usepackage{color}
\usepackage{adjustbox}
\usepackage{color, colortbl}
\definecolor{Gray}{gray}{0.75}
\definecolor{LightCyan}{rgb}{0.50,1,1}

\usepackage[section]{placeins}
\usepackage{amsfonts}
\usepackage{graphicx}
\usepackage{lipsum}
\usepackage{multirow}
\usepackage{booktabs}
\usepackage[table,xcdraw]{xcolor}
\definecolor{cream}{RGB}{222,217,201}

\journal{Acta Materialia}

\begin {document}
\begin{frontmatter}

\title{Effect of substitutional doping and disorder on the phase stability, magnetism, and half-metallicity of Heusler alloys}

\author{N.A. Zarkevich$^{a}$, P. Singh$^{a}$, A.V. Smirnov$^{a}$, Duane D. Johnson$^{a,c}$}
\address{$^a$Ames Laboratory, U.S. Department of Energy, Iowa State University, Ames, Iowa 50011 USA}
\address{$^b$Materials Science \& Engineering, Iowa State University, Ames, Iowa 50011 USA}

\cortext[mycorrespondingauthor]{Corresponding authors email:\textrm{psingh84@ameslab.gov}} 
\fntext[a]{NAZ and PS equally contributed to this work.}

\begin{abstract}
Spintronics is the fast growing field that will play a key role in optimizing power consumption, memory, and processing capabilities of  nanoelectronic devices. Heusler alloys are potential candidates for application in spintronics due to their room temperature (RT) half-metallicity, high Curie temperature, low lattice mismatch with most substrates, and strong control on electronic density of states at Fermi level. In this work, we investigate  the effect of {substitutional doping and disorder} on the half-metallicity, phase stability, and magnetism of Heusler alloys using density functional theory methods. Our study shows that electronic and magnetic properties of half/full-Heusler alloys can be tuned by changing electron-count through controlled variation of chemical compositions of alloying elements. We provide a detailed discussion on the effect of substitutional doping and disorder on the tunability of half-metallic nature of Co$_{2}$MnX and NiMnX based Heusler alloys, where X represents group 13\textendash 16 and period 3\textendash 6  elements of the periodic table. {Based on the idea of electron count and disorder,  we predicted a possible existence of thermodynamically stable half-metallic multicomponent bismuthides, for example, (CuNi$_{3}$)Mn$_{4}$Bi$_{4}$ and (ZnNi$_{7}$)Mn$_{8}$Bi$_{8}$, through substitution doping  at Ni site by specific Cu and Zn  composition in half-Heusler NiMnBi.} We believe that the design guide  {based on electron-counts} presented for half-metals will play a key role in electronic-structure engineering of novel Heusler alloys for spintronic application, which will accelerate the development and synthesis of novel materials.
\end{abstract}


\end{frontmatter}

\bibliographystyle{elsarticle-num}

\section{Introduction}

{\par}
Heusler alloys \cite{Heusler1903} are a large class of naturally occurring \cite{JSSChem43p354y1982} and manufactured \cite{HeuslerAlloys2016} ternary intermetallic compounds with L2$_{1}$ full Heusler (fH) or C1$_{b}$ half-Heusler (hH) structure (sometimes called semi-Heusler), which can be partially disordered  \cite{SciRep8n1p9147y2018,ActaMat142p49y2018,ActaMat148p216y2018,ActaMat115p308y2016,ActaMat107p1359y2016,ActaMat104p210y2016,PRB94p024418y2016}. Many Heuslers are ferromagnetic (FM), with Curie temperatures $T_c$ between 200 and 1100 K \cite{Buschow1983,PJW1971,Brown2000}. The ones with $T_c$ near room temperature were considered for magnetocalorics \cite{JPhysD51n2p024002y2018,JPhysD49n39p395001y2016}. Some alloys have  a band gap in the minority spins, which may (as in half-metals \cite{Groot1983,Hanssen1986,JPhysCM1p2341y1989,Fujii1989,Fujii1990,Fuji1995,Galanakis2002,Galanakis2002_1})  or may not be at the Fermi energy.  A half-metal conducts electrons with one spin orientation, but acts as an insulator or a semiconductor (with a band gap at the Fermi energy $E_F$) to those with the opposite spins.  Half-metals are  used in small-scale magnetic and spin filtering devices and switches  \cite{NatureComm5p3682y2014}, and find applications in magnetic materials \cite{Krishnan2016}. The band-gap engineering \cite{Science235n4785p172y1987,Science352n6292p1446y2016} helps to create materials for advanced electronic devices. Such solid-state devices are used in spintronics -- a rapidly developing branch of electronics, incorporating electron spin degree of freedom \cite{Science294p1488y2001,Spintronics2013}. Materials discovery can be greatly accelerated by theoretical guidance  \cite{JPhysD51n2p024002y2018,NRM3n5p5y2018,NRM2p17053y2017,NRM1p15004y2016,JMS47n21p7317y2012},  in particular, electronic-structure engineering of doped and chemically disordered alloys. However, there has been scant attention on disorder based Heusler alloys. Composition tuning, i.e., disorder, in these alloys can modify the Fermi level $E_{F}$, and alter the electronic, magnetic, and other physical properties \cite{Galanakis2002_1,Hanssen1990,Raphael2002,Complexity11p36y2006}.

Here, we study the electronic structure and its dependence on composition in fully and partially ordered Heusler alloys using density-functional theory (DFT) methods, some of which are half-metallic.  The electronic properties of the doped fH M$_{2}$MnX and hH MMnX alloys with M=\{Co, Ni\} and X=\{Al, Ga, In, Tl; Si, Ge, Sn, Pb; P, As, Sb, Bi; S, Se, Te\}, elements from groups 13\textendash 16 and periods 3\textendash 6 in the Mendeleev Periodic Table were analyzed in detail. We also mix Mn with Fe in Co$_2$(Mn$_{1-x}$Fe$_{x}$)A, where A is Sn or Sb, and $0 \! \le \! x \! \le \! 1$. We show that in fH Co$_{2}$MnX alloys, the width of the minority-spin band gap $E_{gap}$ decreases along the period (3\textendash 6) and increases along the group (13\textendash 16) of X. 

We found that the electronic structure analysis of fH Co$_{2}$MnX and hH NiMnX shows compositional interval for half-metals in terms of the {\it electron count} ($n$). The considered hH NiMnX alloys show a energy gap in the minority spins at (band-gap) or slightly above (pseudo-gap) the Fermi level. Five of them (with X=Si, Ge;  P, As, Sb) could be half-metals, but only one (NiMnSb) was found stable. The structural stability restricts design of hH half-metals. Based on electron count per atom analysis, we predicted a possible existence of weakly stable half-metallic multicomponent bismuthides.  Half-metals often have lower formation energies (and hence are more stable) than metals. Considering a doped hH NiMnBi, we found that adding from 0.3 to 1.3 electrons per formula unit ($e^{-}$/f.u.) would turn this compound into a half-metal; we confirm this prediction for the quaternary CuNi$_{3}$Mn$_{4}$Bi$_{4}$ and ZnNi$_{7}$Mn$_{8}$Bi$_{8}$ alloys.  The  stability of Heusler alloys was analyzed by considering relative structural energies and cross-sections of the ground-state hull. The phase stability analysis suggests that most of these alloys are energetically stable. Small formation enthalpies calculated from DFT methods indicates the preference of the stable alloys for substitutional disordering. Our results provide a valuable guidance for adjustment of the width and position of the band gap in the minority spins relative to the Fermi energy $E_{F}$.

\section{\label{Methods}Computational Methods}

\subsection{Density-functional theory methods}

\noindent
{\bf Vienna Ab-initio Simulation Package (VASP):~} We used VASP \cite{VASP1,VASP2} with the projector augmented waves (PAW) \cite{PAW,PAW2} and Perdew, Burke and Ernzerhof revised for solids (PBEsol) exchange-correlation functional \cite{PBESol2008}. We used a dense $\Gamma$-centered mesh \cite{MonkhorstPack1976} with at least 72 $k$-points per {\angstrom}$^{-1}$ for the Brillouin zone integration. The tetrahedron method with Bl\"ochl corrections \cite{PRB62p6158} was used to calculate electronic density of states (DOS), while Gaussian smearing with $\sigma=0.05\,$eV was used within the conjugate gradient algorithm for the full structural relaxation at zero pressure. We calculated the structural energies, using either primitive unit cells for ternary line compounds, or supercells for the multi\-component alloys. For example, CuNi$_{3}$Mn$_{4}$Bi$_{4}$ was considered in a 12-atom conventional cubic unit cell of the hH structure, while ZnNi$_{7}$Mn$_{8}$Bi$_{8}$ was addressed in a 24-atom $6.141\times6.141\times12.305\,${\angstrom} supercell using $12\times12\times6$ $k$-point mesh. Co$_2$Mn$_1$(Sn$_n$Sb$_{1-n}$) alloys were considered using a decorated 16-atom conventional cubic unit cell of the fH structure, shown in Fig.~\ref{fig1str}a. {The phonon calculations were performed using supercell (96 atoms) of the conventional NiMnBi $C1b$ unit cell (12 atoms), and finite displacement approach with atomic displacement distance of 0.01 \AA~\cite{phonopy}.}

\noindent
{\bf Korringa-Kohn-Rostoker coherent potential approximation (KKR-CPA):~} Green's function based KKR-CPA method  \cite{KKR1,KKR2}\cite{MECCA} is suitable for both fully ordered and partially disordered multicomponent systems. It treats multiple scattering in the atomic spheres approximation (ASA) \cite{AJ2009}. We use the optimal local basis set \cite{MECCA} with the radii of scattering spheres, determined from the atomic charge distributions around each atom and its neighbors.  The periodic boundary correction takes into account both electrostatic energy \cite{TotEnCorr_Christensen1985} and Coulomb potential. Use of the variational definition \cite{AJ2012} of the muffin-tin potential zero \emph{$\text{v}_{0}$} and a proper representation of the topology of charge density in the optimal basis set allows to approach accuracy of the full-potential methods \cite{PRB90p205102y2014}. 

Our truncated optimal basis set with $l_{\max}$=3 includes {\emph{s-, p-, d-}} and {\emph{f-}}orbitals; we find a negligible sensitivity of energy differences to the higher-order spherical harmonics. Integration in a complex energy plane uses the Chebyshev quadrature semicircular contour with 20 points. The Brillouin zone integration is performed using a special $k$-points method \cite{MonkhorstPack1976} with a $12\times12\times12$ mesh for disordered or ordered systems with 4 atoms per unit cell and $8\times8\times8$ $(6\times6\times6)$ mesh for ordered systems with 16 atoms/cell.  We use the Perdew, Burke and Ernzerhof revised for solids (PBEsol) \cite{PBESol2008}.  Self-consistency in spin-polarized calculations were was achieved using the modified Broyden's second method \cite{Broyden1988}. The homogeneous substitutional chemical disorder is considered using the coherent potential approximation (CPA) \cite{JohnsonCPA} with the mean-field corrections \cite{JP1993}. 

\noindent
{\bf Tight-binding-Linearized Muffin-tin Orbital Method:~}We use the van Leeuwen-Baerends corrected LDA exchange (LDA+vLB) \cite{VLB, PS2013,PS2015book,PRB93p085204y2016,JPCM2017Singh,PhysRevB.96.054203Singh,DATTA2020125} with the exchange-correlation energy parameterization \cite{vBH}. An improved ASA basis of TB-LMTO \cite{TBLMTO} includes atomic spheres and empty spheres (ESs). The sum of the volumes of all the spheres is equal the total volume of the periodic unit cell.  The radii of the spheres are chosen to have their overlaps close to the  local  maxima or saddle points  of  the  electrostatic potential.  The positions of the ES centers are at the high-symmetry points between atoms. The relaxed atomic positions are taken from VASP output.  ESs are treated as empty sites with no cores and small electronic charge density. Because the contribution of exchange in empty spheres is very small, the vLB-correction is calculated only in the atomic spheres  \cite{VLB, PS2013,PS2015book,PRB93p085204y2016,JPCM2017Singh,PhysRevB.96.054203Singh,HSP,HSP1,HSH2014}. 

Self-consistently was achieved when the change of charge density and energy is below 10$^{-5}$ and $<10^{-4}$ Ry/atom, respectively \cite{TBLMTO}. The $k$-space integration is done using the tetrahedron method with the $12\times12\times12$ mesh. In theory, the Coulomb and exchange self-interactions should be cancelled  \cite{PRB93p085204y2016,JPCM2017Singh,PhysRevB.96.054203Singh}. The LDA and GGA are known to under-estimate band gaps; this problem is less severe in GGA due to the gradient correction \cite{GGA}. On the other hand, the asymptotic vLB correction significantly improves the exchange potential  both near the nucleus  ($r \rightarrow 0$) and at the large-distance limit  ($1/r \to 0$). It correctly describes the valence and conduction-band energies, and provides the band gaps comparable to those measured in experiments  \cite{PRB93p085204y2016, JPCM2017Singh,PhysRevB.96.054203Singh}. 

\subsection{Energy-gap estimate}
The width of the band gap is estimated from the plateau in the total number of the minority-spin electrons $n_{s}(E)$ (more detail in Fig.~\ref{fig4E2}c). Counting electrons in the majority-spin channel, we find the number of electrons n$_{-}$ (n$_{+}$) that should be added [$n \! > \! 0$] or removed [$n \! < \! 0$] to move the highest occupied band (the lowest unoccupied band) in the minority spins to the Fermi level.  These numbers define the compositional interval of half-metallicity in terms of the electron count.

\section{Results and discussion}

\begin{figure}[htp]
\centering
\includegraphics[scale=0.42]{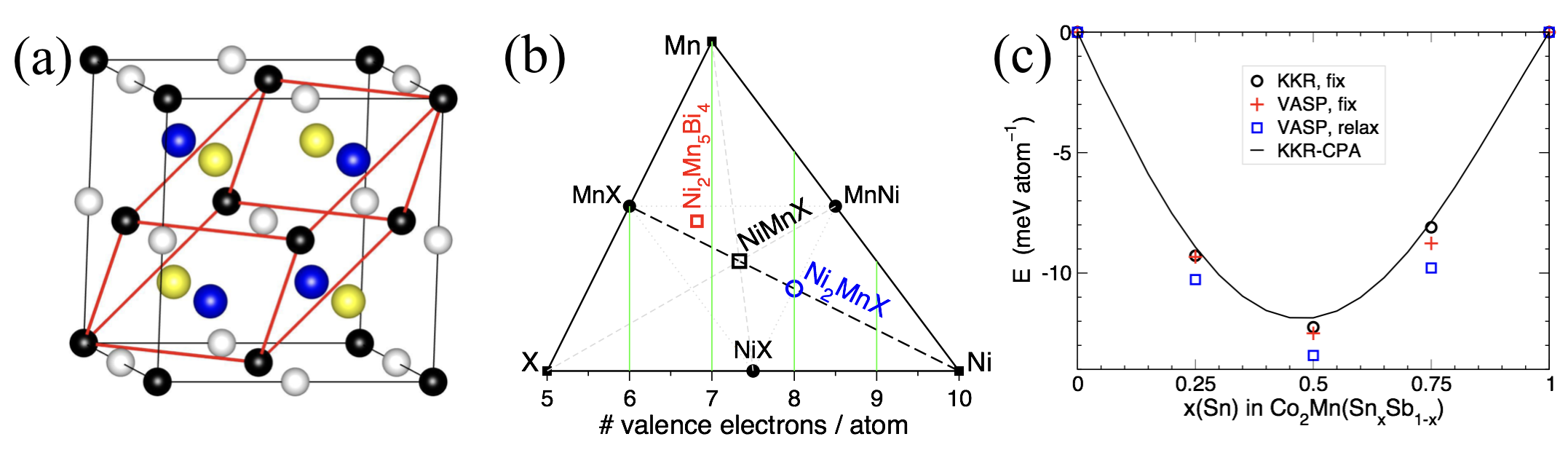}
\caption{\label{fig1str} (a) The quaternary Heusler LiMgPdSn-type cubic structure with $F\bar{4}3m$ space symmetry group (\#216) has atoms on the bcc cites. The C1$_{b}$, cF12 ($F\bar{4}3m$, \#216) hH AgAsMg-type MMnX differs from the L2$_{1}$, cF16 ($Fm\bar{3}m$, \#225) fH Cu$_{2}$MnAl-type M$_{2}$MnX structure by vacancies on $(\frac{1}{2}$,$\frac{1}{2}$,$\frac{1}{2})$ sites (white). M is black (black and white in fH), Mn is blue, and X is yellow. The conventional (cubic) and primitive  unit cells are shown by black and red lines, respectively. (b) Ternary phase diagram with iso-surfaces of electron count per atom (vertical green lines) for Ni-Mn-X alloys with group 15 elements X = \{P, As, Sb, Bi\}. (c) Mixing enthalpies (meV/atom) of Co$_{2}$Mn$_{1}$(Sn$_{x}$Sb$_{1-x}$) at 0 K (DFT).}
\end{figure}

\noindent
{\it Crystal Structure:~} The full-Heusler and half-Heusler ternary alloys, shown in Fig.~\ref{fig1str}a, crystallize in L2$_{1}$ and C1$_{b}$ structures, respectively.~The fH M$_{2}$ZX structure (L2$_{1}$, cF16, with Cu$_{2}$MnAl prototype, $Fm\bar{3}m$ cubic space group \#225) has a cubic conventional unit cell with 16 sites of 3 types, which consists of four interpenetrating fcc sub\-lattices with positions (0,0,0) and ($\frac{1}{2}$,$\frac{1}{2}$,$\frac{1}{2}$) for M, ($\frac{1}{4}$,$\frac{1}{4}$,$\frac{1}{4}$) for Z (Mn) and ($\frac{3}{4}$,$\frac{3}{4}$,$\frac{3}{4}$) for X atoms. The primitive unit cell has only 4 sites. The atoms decorate a body centered cubic (bcc) lattice. The hH M$_{1}$ZX structure (C1$_{b}$, cF12, with MgAgAs prototype, $F\bar{4}3m$ space group \#216) differs from cF16 fH by a vacancy on ($\frac{1}{2}$,$\frac{1}{2}$,$\frac{1}{2}$) sites. 

\noindent
{\it Substitutional elements:~} We consider M$_{2}$ZX fH and M$_{1}$ZX hH alloys with M=\{Co, Ni\}; Z=\{Mn\}; and X=\{Al, Ga, In, Tl; Si, Ge, Sn, Pb; P, As, Sb, Bi; S, Se, Te\}, as well as mixed Co$_2$(Mn$_{1-x}$Fe$_{x}$)A with A=\{Sn or Sb\}. Our investigation covers both fully ordered and partially disordered (substitutionally doped) alloys.

\noindent
{\it {Electron count and composition}:}~Historically, the search for half-metals started from line compounds \cite{NatComm5p3974y2014}. We emphasize that  the Fermi level $E_F$ remains in the band gap within a range of energies, bounded by the highest occupied and the lowest unoccupied states in the minority spins, and there is a corresponding range of half-metallic compositions. Indeed, the number of valence electrons $N (c)$ depends on elemental composition $c$.  Fig.~\ref{fig1str}b illustrates the inverse, many-to-one $c(N)$ relation for the ternary Ni-Mn-X alloys. Thus, a fixed $N$ specifies a set of compositions,  leaving a sufficient freedom in choosing chemical elements for doping. For example,  hH alloys with compositions NiMn(P$_x$As$_y$Sb$_z$Bi$_w$), where $0 \le \{x,y,z,w \} \le 1$ and $x+y+z+w=1$, have $N \! = \! 7\frac{1}{3}$, see Fig.~\ref{fig1str}b. Below we express the electron count $n$ as a difference in the number of valence electrons relative to a line compound.

\noindent
{\it {\label{GuidanceRange}}Half-metallic range:}~In each compound,  we find a band gap in the minority spins (if any) of the width $E_{gap} = (E_+ - E_-)$ at the electronic energies $E$ in the range $E_- \! \le \! ( E - E_F ) \! \le \! E_+$.  A compound is  half-metallic, if the Fermi level happens to be in this gap (thus, $E_- \! < \! 0$ and $E_+ \! > \! 0$). 

{\par} 
The number of electronic states at energies below $E_0$ is 
\begin{equation} 
\label{eqN}
N (E_0) = \int_{-\infty}^{E_0} g(E ) dE ,
\end{equation}
where $g$ is the electronic density of states (DOS) for both spins. At zero electronic temperature, all states below the Fermi energy $E_F$ are filled, hence $N(E_F)$ is equal to the total number of electrons $N$.  (If one uses pseudo\-potentials or does not count core electrons, then one can use the total number of valence electrons.) A band gap in the minority spins  covers a ragne of both occupied and empty electronic states  from $ N_- \equiv N(E_F + E_-)$ to $ N_+ \equiv N(E_F + E_+)$;  those are the edges of the band gap.  Next, we use eq.~\ref{eqN} to find the differences 
\begin{equation}
 \label{eqndef}
 n_{\pm} \equiv (N_{\pm} -N) = \int_{E_F}^{E_F + E_\pm} g(E ) dE .
\end{equation}
The signs of $E_\pm$ and $n_\pm$ coincide, because $g \ge 0$. A compound is  half-metallic, if $n_- <0$ and $n_+ >0$, see Table~\ref{t2n} \& \ref{t4n}.  

\begin{table}[htp]
\centering
\caption{\label{t2n}  
The minority-spin band gap width $E_{gap}$ (eV) and position in terms of energies $E_{\pm}$ (relative to $E_F$) and electron count $n_{\pm}$ 
 in full-Heusler Co$_{2}$MnX ternary line compounds.  The unstable compounds are marked with an asterisk ($^{*}$).}
\begin{tabular}{l|rl|rl|ll}
\hline 
\multirow{2}{*}{ X} & $n_{-}$  & $n_{+}$  & $E_{-}$  & $E_{+}$  & $E_{gap}^{GGA}$  & $E_{gap}^{vLB}$ \tabularnewline
 & \multicolumn{2}{c|}{($e$/f.u.)} & \multicolumn{2}{c|}{(eV)} &  \multicolumn{2}{c|}{(eV)} 
 \tabularnewline
\hline 
Al  & 0.07  & 0.84  & 0.06  & 0.62  & 0.56  & 0.38 \tabularnewline
Ga  & 0.5  & 0.84  & 0.24  & 0.55  & 0.31  & 0.23 \tabularnewline
In$^{*}$  & 0.9  & 1.0  & 0.54  & 0.64  & 0.10  & \textendash{} \tabularnewline
Tl$^{*}$  & 1.1  & 1.1  & 0.63  & 0.63  & 0.0  & 0.0 \tabularnewline
\hline 
Si  & -0.6  & 0.35  & -0.39  & 0.34  & 0.73 & 0.49 \tabularnewline
Ge  & -0.2  & 0.4  & -0.12  & 0.38  & 0.50  & 0.37 \tabularnewline
Sn  & 0.1  & 0.45  & 0.09  & 0.44  & 0.35  & 0.37 \tabularnewline
Pb$^{*}$  & 0.35  & 0.5  & 0.28  & 0.43  & 0.15  & 0.11 \tabularnewline
\hline 
P$^{*}$  & -1.1  & 0.05  & -0.76  & 0.04  & 0.80  & 0.55 \tabularnewline
As$^{*}$  & -0.6  & 0.05  & -0.43  & 0.05  & 0.48  & 0.44 \tabularnewline
Sb  & -0.5  & 0.03  & -0.51  & 0.02  & 0.53 & 0.38 \tabularnewline
Bi$^{*}$  & -0.1  & 0.2  & -0.08  & 0.27  & 0.35  & 0.26 \tabularnewline
\hline 
\end{tabular}
\end{table}

Within the frozen-band approximation, let us increase the total number of electrons from $N$ to $(N+n)$, where negative $n$ stands for subtracting electrons. The added or subtracted electrons will shift $E_F$ to the band gap (thus making the system half-metallic), if $ {n_- \le n \le  n_+}$. In general, doping a conductor with electronic donors or acceptors changes the number of electrons, shifts the Fermi level, and adds impurity states.  If those impurity states are not in the band gap,  then half-metallicity of the doped system is approximated by  inequality (${n_- \le n \le  n_+}$).  Lattice relaxations around impurities and defects broaden the bands and narrow the band gap. In addition, sometimes electronic bands of the dopant can appear in the band gap. These effects narrow both the band gap and the half-metallic compositional range. With this caution, inequality (${n_- \le n \le  n_+}$) and  Tables~\ref{t2n} and  \ref{t4n} provide a practical guidance for engineering of the half-metallic Heusler alloys.

\begin{table}[htp]
\centering
\caption{\label{t4n} 
The minority-spin band gap $E_{gap}$ and its position in terms of energies $E_{\pm}$ and electron count $n_{\pm}$ in half-Heusler NiMnX compounds.}
\begin{tabular}{l|rl|rl|ll}
\hline 
\multirow{2}{*}{ X}  & $n_{-}$  & $n_{+}$  & $E_{-}$  & $E_{+}$  & $E_{gap}^{GGA}$  & $E_{gap}^{vLB}$ \tabularnewline
 & \multicolumn{2}{c}{($e$/f.u.)} &  \multicolumn{2}{c|}{(eV)} &  \multicolumn{2}{c|}{(eV)}   
 \tabularnewline
\hline 
Al$^{*}$  & 0.6  & 1.2  & 0.10  & 0.90 & 0.80  & $-$ \tabularnewline
Ga$^{*}$  & 0.9  & 1.3  & 0.30  & 1.00 & 0.70  & $-$ \tabularnewline
In$^{*}$  & 1.25  & 1.5  & 0.46  & 0.88  & 0.40  & 0.32 \tabularnewline
Tl$^{*}$  & 1.4  & 1.6  & 0.60  & 0.90  & 0.30  & 0.27 \tabularnewline
\hline 
Si$^{*}$  & -0.37  & 0.45  & -0.35  & 0.56  & 0.91  & 0.8 \tabularnewline
Ge$^{*}$  & -0.0  & 0.57  & -0.0  & 0.7  & 0.70 & 0.6 \tabularnewline
Sn$^{*}$ & 0.40  & 0.72  & 0.28  & 0.74  & 0.46  & 0.43 \tabularnewline
Pb$^{*}$  & 0.65  & 0.87  & 0.45  & 0.79  & 0.34  & 0.28 \tabularnewline
\hline 
P$^{*}$  & -0.67  & 0.22  & -0.66  & 0.28  & 0.93  & 0.7 \tabularnewline
As$^{*}$  & -0.37  & 0.30  & -0.38  & 0.39  & 0.77  & 0.6 \tabularnewline
Sb  & -0.17  & 0.22  & -0.27  & 0.28  & 0.55  & 0.45 \tabularnewline
Bi$^{*}$  & 0.06  & 0.31  & 0.07  & 0.47  & 0.40  & 0.34 \tabularnewline
\hline 
\end{tabular}
\end{table}

Known half-metals among ternary systems include full Heusler Co$_{2}$MnX with X=\{Si, Ge, Sb\} and half-Heusler NiMnSb compounds, {which is in agreement with the findings of Ma \etal \cite{PhysRevB.95.024411}.} We also emphasize that not only the line compounds, but also ranges of compositions (with off-stoichiometric disorder) can be half-metallic. We compute the compositional ranges, which keep known half-metals half-metallic, and predict the level of doping, which can turn others (like Co$_{2}$MnSn and hypothetical NiMnBi) into half-metals. Table~\ref{t4n} can be used to design half-metals of composition (Ni$_{1-z-y}$Cu$_{z}$Zn$_{y}$)MnX with X=\{Al, Ga, In, Tl; Si, Ge, Sn, Pb; P, As, Sb, Bi\}. For each X, we predicted a range of $n=(z+2y)$, at which this doped half-Heusler alloy is half-metallic. For negative $n$, Ni can be mixed with Co and Fe to make (Ni$_{1-z-y}$Co$_{z}$Fe$_{y}$)MnX alloys, where $(z+2y)=|n|>0$. Dopants narrow the band gap, hence the actual composition range could be narrower than that predicted from electron count.

\noindent
{\it Phase stability $-$ formation energy of Co-Mn-Sn-Sb Heusler alloys:}~ We show DFT calculated mixing enthalpies (E$_{mix}$) of Co$_{2}$Mn$_{1}$(Sn$_{x}$Sb$_{1-x}$) with fixed direct atomic coordinates (fix) in Fig.~\ref{fig1str}c, and compared those with fully relaxed atoms (relax). The ordered structures with 16-atom cells are addressed by VASP and KKR. On the other hand, the disorder on the Sn/Sb sublattice was modeled in the 4-atom primitive cell (shown in Fig.~\ref{fig1str}a), and E$_{mix}$ was calculated using KKR-CPA. The E$_{mix}$ shows the stability for each structure on the ground-state (GS) hull in Fig.~\ref{fig1str}c. The structures with energies above the GS hull are either unstable or meta\-stable, which might be stabilized by entropy at finite temperature. Our calculations show good agreement for predicted E$_{mix}$ across the DFT methods, which emphasizes the accuracy of DFT calculations. 

To avoid a systematic error in the energy of $\alpha$-Mn,  resulting from application of DFT to an astonishingly complex $\alpha$-Mn crystal structure with non-collinear moments,  we use the semi-empirical energy of metallic Mn, obtained from the calculated energies of metallic Bi \cite{Complexity11p36y2006} and  ferrimagnetic MnBi (with NiAs structure) \cite{APLMat2p032103y2014}, and the experimental \cite{Shchukarev1961} E$_{mix}$ of MnBi of $-0.204 \,$eV/f.u. By construction, the calculated formation energy of MnBi coincides with this experimental value.

\begin{figure*}[htp]
\centering
\includegraphics[scale=0.4]{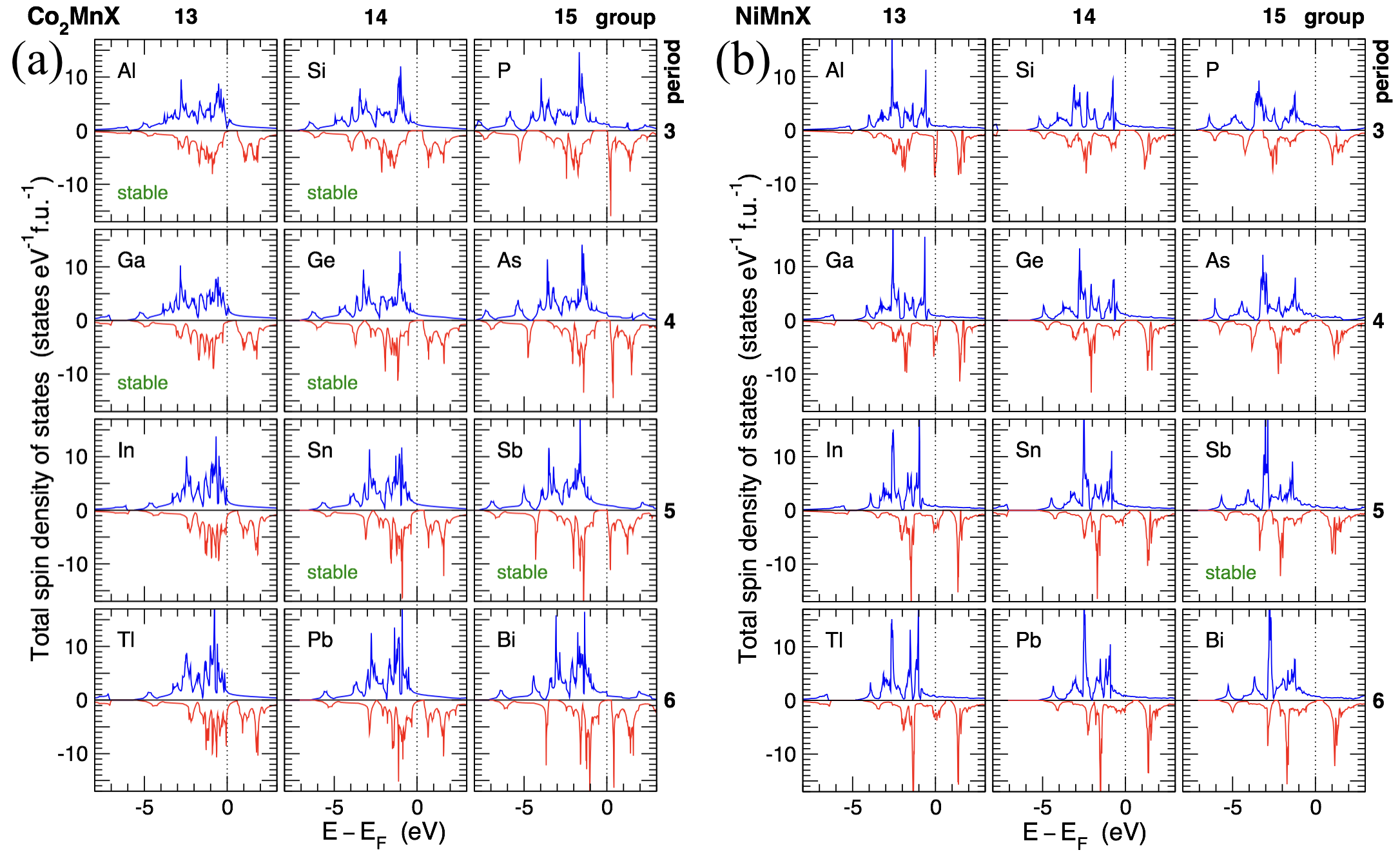} 
\caption{\label{fig5dos} Total spin DOS (states/eV$\cdot$f.u.) of the (a) fH Co$_{2}$MnX, and (b) hH NiMnX alloys  for 12 elements X (group 13-15, period 3-6). Stable compounds are known from experiment \cite{Buschow1983,PJW1971,JAP105n1p013716y2009}. }
\end{figure*}

\noindent
{\it Electronic-structure engineering:~} We compare the calculated electronic DOS for the ternary line compounds Co$_{2}$MnX and Ni$_{1}$MnX in Fig.~\ref{fig5dos}. We use experimental data to mark the known stable compounds. Our values of $E_{\pm}$ and $n_{\pm}$, defined in section IIIA, are provided in Tables I and II for fH-Co$_{2}$MnX and hH-NiMnX (which have the minority-spin band gap in the vicinity of E$_{F}$). This data is useful for engineering materials with a modified electronic structure and for predicting the band gaps in disordered alloys, some of which will be considered in section IV B.

In Fig.~\ref{fig5dos}a,b, we plot total spin DOS (states/eV$\cdot$f.u.) for  (a) fH Co$_{2}$MnX, and (b) hH NiMnX alloys. Quite interestingly, we found energy gap or pseudo-gaps in the minority-spin near/at $E_F$ for both stable and unstable ferromagnetic compounds. In fH Co$_2$MnX the gaps are at or near $E_F$. Similarly, in hH NiMnX they are at or slightly above $E_F$. In contrast, there are no half-metals among fH Ni$_2$MnX alloys (see Fig.~\ref{DOS_Ni2MnX_fH}). Because Ni has more electrons than Co, the difference between the fH Co$_2$MnX and Ni$_2$MnX is expected from the electron count, as shown in Table~\ref{t2n}~\&~\ref{t4n}. Indeed, the gaps are near $E_F$ in Co$_2$MnX, but quite far ($\sim \! 1\,$eV) below $E_F$ in Ni$_2$MnX.

Tables~\ref{t2n} and \ref{t4n} show the  band gap $E_{gap} = (E_+ - E_-)$ for the minority spins  from GGA (VASP) and LDA+vLB \cite{Singh2019} calculations. The gap extends from $E_-$ to $E_+$ (both energies are relative to $E_{F}$); for half-metals, these are energies of the highest occupied and the lowest unoccupied bands in the minority spins. The line compound was categorized half-metallic, if $E_{-}\le 0$ and $E_{+}\ge 0$ (consequently, $n_{-}\le 0$ and $n_{+}\ge 0$).  For half-metals with off-stoichiometric disorder, the level of doping in terms of $n$ (added electrons per formula unit) lies within the range $[n_{-}, n_{+}]$. As explained in section~\ref{GuidanceRange}, dopants narrow the gap, hence the actual compositional range can be narrower. 

The LDA/GGA functionals usually underestimate the band gap in semiconductors and half-metals. This systematic error narrows the predicted compositional range for half-metals.  An error in the predicted band gaps is larger in LDA, smaller in GGA \cite{GGA}, and is expected to cancel for the exact exchange and correlation. The LDA+vLB used in this work is a semi-local approximation for the exact exchange that was found to provide quite accurate predictions of semiconductor band-gaps \cite{PRB93p085204y2016}. The LDA correlation with modified LDA+vLB (exact) exchange was used as shown in Tables~\ref{t2n} and \ref{t4n}.


\subsection{Alloys with substitutional atomic disorder and comparison with experiments}
In materials engineering, consideration of multi\-component alloys with a partial  atomic disorder can be challenging.  We analyze quantitatively the selected fH alloys in section~\ref{PDA}. A fH alloy has three sublattices (Fig.~\ref{fig1str}a). A substitutional doping is possible on each sublattice. In section~\ref{PDA}, we consider several mixed fH alloys (with disorder on one of the sublattices) and validate predictions, based on the electron count. 

Addressing the ternary line compounds, we fully relaxed each structure. We found a reasonable agreement between the calculated and measured \cite{Buschow1983,PJW1971} lattice constants $a$ and magnetic moments $M$, see Tables~\ref{t2c} and Table~\ref{t3a}. We considered both ferromagnetic (FM) and antiferromagnetic (AFM) spin ordering. For the Co$_{2}$MnX systems in Fig.~\ref{fig5dos}, we find that FM ordering is preferred, in agreement with the previous calculations \cite{PRB28p1745y1983}. The calculated \cite{VASP1,VASP2} magnetization $M$ of Co$_{2}$MnSn is $5.03\,(5.04)\,\mu_{B}$ per formula unit  with Sn d-electrons in the core (valence) at the equilibrium lattice constant $a_{0}=5.9854\,${\angstrom}.  It reasonably agrees with the experimental \cite{Webster1969} $M$ of $5.08\,\mu_{B}$. Earlier full-potential FLAPW \cite{FLAPW2002_Picozzi} and our current KKR calculations provide   $5.0\,\mu_{B}$/f.u. at $a_{0}=5.964\,${{\angstrom} and $4.98\,\mu_{B}$/f.u. at $a_{0}=5.995\,${\angstrom}, respectively. However, there is an experimental uncertainty in the weight of the powder sample, only a fraction of which is the desired phase \cite{Buschow1983}.  The imprecisely measured magnetization of the Co$_{2}$MnSn and Co$_{2}$MnSb  polycrystalline powder  samples \cite{Buschow1983} is  below the expected theoretical value, which must be integer for a half-metal,  and Co$_{2}$MnSb was found a half-metal in experiments \cite{Buschow1983}.

\subsection{NiMnX half-Heusler alloys }

\begin{figure}[htp]
\centering
\includegraphics[scale=0.35]{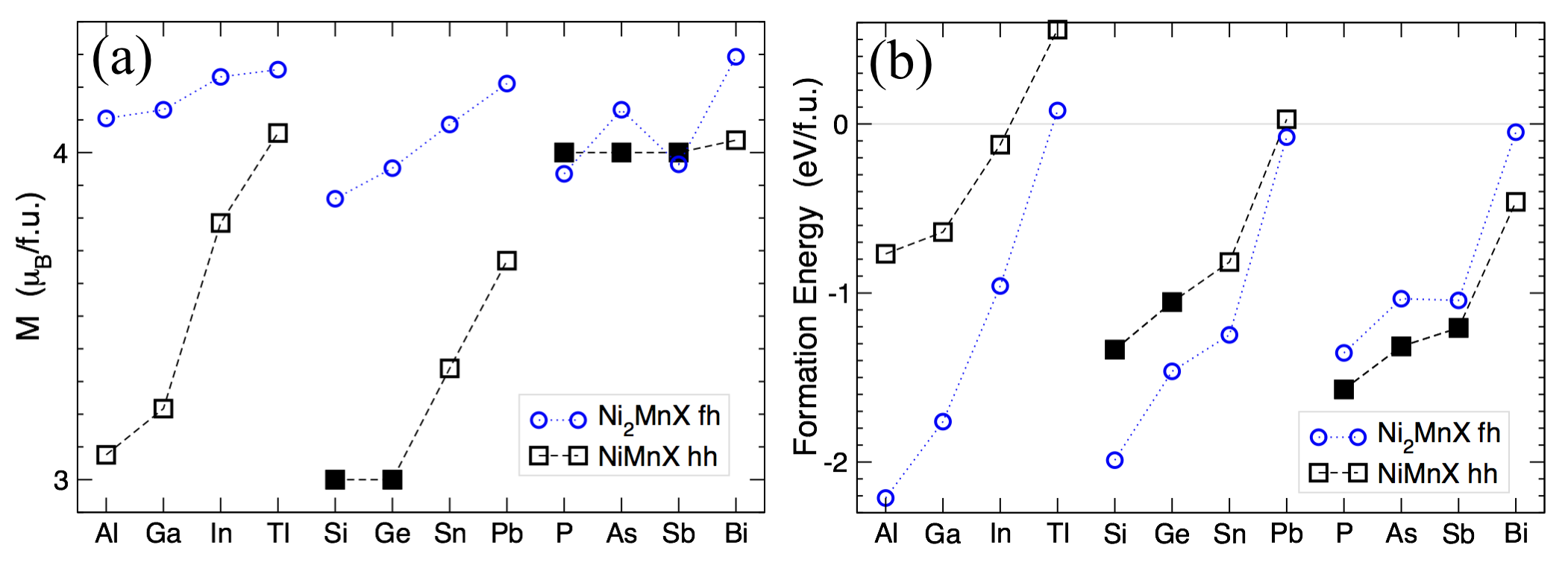} 
\caption{\label{fig2M} (a) Calculated magnetization (Bohr magnetons per formula unit) for hH NiMnX (black squares). Half-metals are solid symbols. The fH Ni$_{2}$MnX (blue circles) are ferrimagnetic in nature (see fH detail in supplement).(b) Formation energy (electron-volts per formula unit) of the hH NiMnX (black squares). Half-metals are solid symbols. We provide formation energy of fH Ni$_{2}$MnX alloys (blue circles) for comparison (see fH detail in supplement).}
\end{figure}

Among the hH~NiMnX alloys, where X is one of \{Al, Ga, In, Tl; Si, Ge, Sn, Pb; P, As, Sb, Bi\},  NiMnSb is the only known stable ternary compound,  which had been claimed to be a half-metal \cite{Groot1983,NatPhys12p855y2016},  but its half-metallicity was questioned by some measurements \cite{PRB68p104430y2003}  and calculations \cite{PRB81p054422y2010}. We calculate magnetization (Fig.~\ref{fig2M}a), formation energy (Fig.~\ref{fig2M}b), and electronic DOS (Fig.~\ref{fig5dos} and Fig.~\ref{DOS_Ni2MnX_fH}) of fH~Ni$_{2}$MnX and hH~NiMnX alloys. Magnetization of the half-metallic hH alloys increases from 3 $\mu_{B}$/f.u. for NiMnSi and NiMnGe containing group 14 elements to 4 $\mu_{B}$/f.u. in group 15. An anti-ferromagnetic (AFM) spin ordering was found preferable for most compounds of sulfur, a group 16 element with a small atomic size and high electronegativity.

\begin{figure}[htp]
\centering
\includegraphics[scale=0.35]{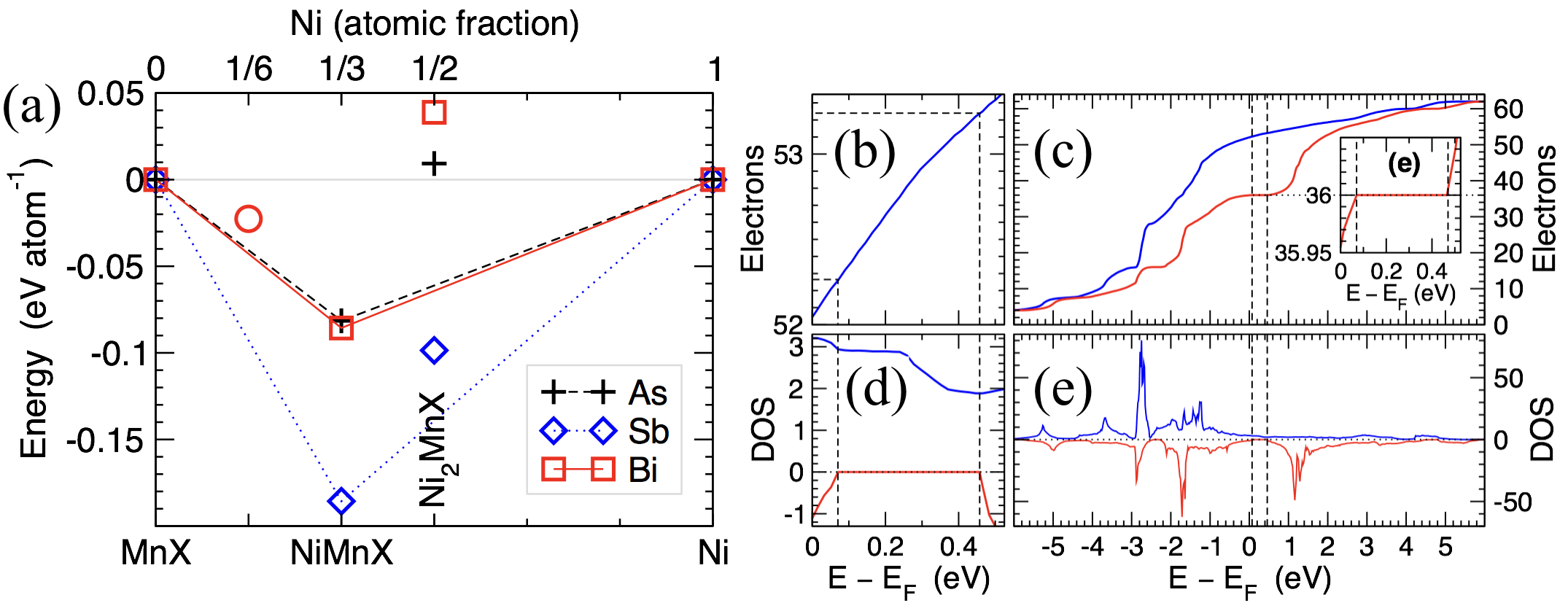} 
\caption{\label{fig4E2} (a) Energy (eV/atom) of the hH NiMnX compounds relative to that of metallic fcc Ni and MnX (X = As, Sb, Bi), see  Fig.~\ref{fig1str}b. Red circle is the energy of the phase-segregated Ni$_{2}$Mn$_{5}$Bi$_{4}$ and metallic Bi. Lines are cross-sections of the ground-state hull. (b-d) Number of electrons (per unit cell) and DOS (states/eV cell) of the hH NiMnBi  alloy}
\end{figure}

Among the hH alloys, we found five half-metals: they are NiMnX with X=\{Si, Ge, P, As, Sb\}. The first two with a  group 14 element (NiMnSi, NiMnGe) have magnetization of 3 $\mu_{B}$/f.u., which increases to 4 $\mu_{B}$/f.u. for the last three (NiMnP, NiMnAs, NiMnSb) with X from group 15. In spite of having large magnetic moments, cubic systems can not make hard magnets due to absence of anisotropy;  a structural anisotropy is necessary for a magnetic anisotropy in a hard magnet. All considered hH NiMnX alloys have a gap in the minority spins at or slightly above the Fermi level (Fig.~\ref{fig5dos} and Table~\ref{t2n}). In general, a peak with a maximum in electronic DOS at the $E_{F}$ destabilizes the alloy. Electron (or hole) doping is one option, which can help in shifting $E_{F}$ away from this peak, thus reducing $g(E_{F})$. Half-metals are the most stable compounds among the considered hH alloys (Fig.~\ref{fig2M}b). Fig.~\ref{fig4E2}a shows relative energies of the line compounds.  The fH Ni$_2$MnX alloys with X=\{As, Bi\} are not stable, because their energies are well above the ground-state hull.  Although the fH Ni$_2$MnSb could be stabilized by entropy at finite temperature $T$, we predict that at low $T$ it tends to segregate towards Ni$_{2 - \delta }$MnSb and metallic Ni, although diffusion is limited at low $T$. The hH NiMnSb is known to be stable in experiment, and Fig.~\ref{fig4E2}a does not question its stability. 

NiMnBi is only 0.06 $e^{-}$/f.u. away from being a half-metal;  it  has a gap in the minority spin channel above the $E_{F}$ (Fig.~\ref{fig4E2}b-d), and might be turned into a half-metal by electron doping,  such as a partial substitution of Ni by Cu or Zn in Fig.~\ref{fig7dos4}. From the estimate of the needed level of doping (see Fig.~\ref{fig4E2}b-d and Table~\ref{t2n}), we predict that CuNi$_{3}$Mn$_{4}$Bi$_{4}$ and ZnNi$_{7}$Mn$_{8}$Bi$_{8}$ should be half-metals. The total DOS plot for hH NiMnBi compared to for fH Ni$_{2}$MnBi in Fig.~\ref{fig7dos4} shows an opening in the energy-gap in the minority spin, however, some electronic states are still present at the $E_F$. To fill those electronic states at  $E_F$, we electron doped Ni by adding Cu (0.25)/Zn(0.125). Our supercell calculations show a clear band-gap in the minority spins, although doping narrowed the band gap (see Table~\ref{t4n}).

\begin{figure}[htp]
\centering
\includegraphics[scale=0.38]{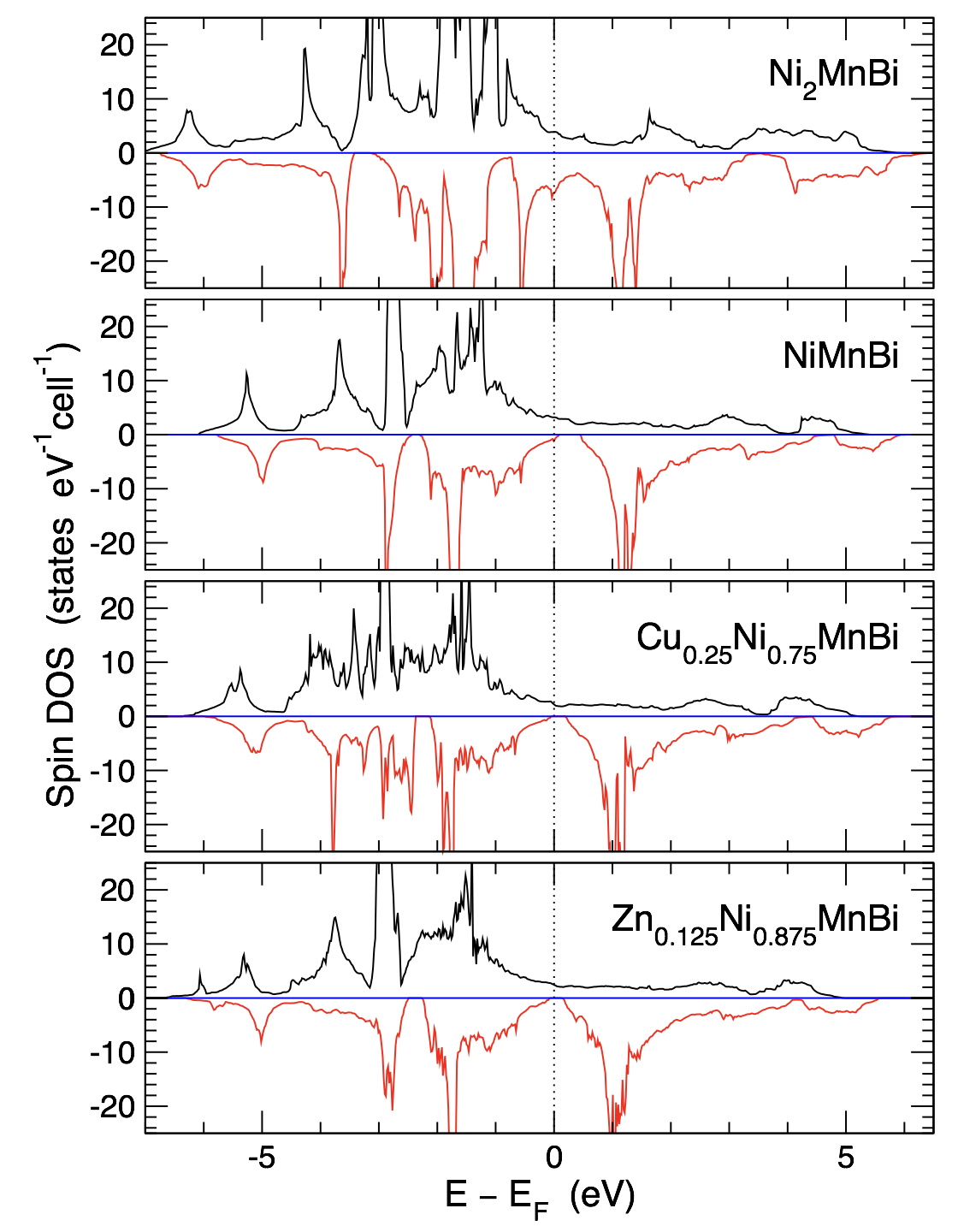} 
\caption{\label{fig7dos4} Total spin DOS (states/eV-f.u.) of the fH Ni$_{2}$MnBi (unstable), hH NiMnBi, and doped hH alloys with partially substituted Ni. Doping shifts the Fermi level and narrows the band gap in the minority spins.}
\end{figure}

Half-metals, in general,  are thermodynamically more stable than other metallic half-Heuslers as shown in Fig.~\ref{fig2M}b. We want to emphasize that a negative formation energy (Fig.~\ref{fig2M}b) relative to the elemental ground states does not necessarily indicate stability of a given phase, because there could be other (more stable) phases nearby, such as the weakly stable Ni$_{2}$Mn$_{5}$Bi$_{4}$, NiBi and MnBi in Ni-Mn-Bi system (see Fig.~\ref{fig1str}b and Table~\ref{t1E}). Energies of all possible competing structures are needed for constructing the complete ground-state hull. While an alloy with a negative formation energy might be stable, a positive energy relative to any set of phases indicates that a structure is above the ground-state hull and is definitely unstable towards segregation at low $T$. However, some of the metastable structures are technologically feasible, especially if the phase transformation or segregation has high enthalpy barriers or a metastable phase is thermally stabilized by entropy; examples are iron steels \cite{Bugayev1971,PRB91p174104,JChemPhys142p064707y2015}, Alnico magnets \cite{Campbell1994}, titanium alloys \cite{PRB93p020104y2016}, shape memory alloys \cite{PRL113p265701y2014,PRB90p060102y2014,C2NEB}, and graphite \cite{JChemPhys38n3p631y1963}. The ferrimagnetic hexagonal MnBi, NiBi, and magnetic cubic Ni$_{2}$Mn$_{5}$Bi$_{4}$ are weakly stable compounds. In particular,  determined from the measured heat of combustion \cite{Shchukarev1961} formation enthalpy of MnBi is only  $-4.7\pm0.1\,$kcal/g-formula ($-0.102\pm0.002\,$eV/atom). The measured formation enthalpy of NiBi is approximately $-0.93\,$kcal/g-formula ($-0.020\,$eV/atom); this value was inaccurate due to inhomogeneity of the sample \cite{Predel1972}. The calculated formation enthalpies of NiBi and Ni$_{2}$Mn$_{5}$Bi$_{4}$ reasonably agree with experiment (Table~\ref{t1E}). {We also estimated the  Curie temperature  (T$_{c}$=$\frac{2}{3}\frac{\Delta{E_{AFM-FM}}}{3k_{B}}$) NiMnBi, CuNi$_{3}$Mn$_{4}$Bi$_{4}$, and ZnNi$_{7}$Mn$_{8}$Bi$_{8}$, which is higher than room-temperature, i.e., 312 K, 324 K, and 310 K, respectively. The above room-temperature T$_{c}$ indicates the possible applicability of half-Heusler compounds in spintronic devices. The prediction of dynamical stability \cite{VILLARREAL2021437} is critical for experimentalists to know if compound can be synthesized. Therefore, we carried out phonon calculations \cite{phonopy} of one of the compound, i.e., NiMnBi see Fig.~\ref{Phonon_NiMnBi}, which shows no imaginary phonon modes. This suggests that the predicted compound is both thermodynamically and dynamically stable, which further establishes the quality of our prediction.}

\begin{table}[htp]
\centering
\caption{\label{t2n2} For selected compounds, $[n_{-},n_{+}]$ range ($e^{-}$/f.u.), $E_{-}$, $E_{+}$, and $E_{gap}=E_{+}-E_{-}$ (eV), and Curie temperature (K) from the supercell calculations, see Table~\ref{t2n} caption.}
\begin{tabular}{ccccccccc}
\hline 
 & $n_{-}$  & $n_{+}$  & $E_{-}$  & $E_{+}$  & $E_{gap}^{GGA}$  & $E_{gap}^{vLB}$ & T$_{c}$ \tabularnewline
\hline 
NiMnBi                                      & 0.06  & 0.31  & 0.07 & 0.47  & 0.40  & 0.34 & 312 \tabularnewline
CuNi$_{3}$Mn$_{4}$Bi$_{4}$  &-0.57  & 1.1 & -0.10 & 0.25 & 0.35 & 0.25 & 324 \tabularnewline
ZnNi$_{7}$Mn$_{8}$Bi$_{8}$  & -0.12 & 0.2  & -0.11 & 0.19 & 0.30 &---& 310 \tabularnewline
\hline 
\end{tabular}
\end{table}

\subsection{Effect of spin-orbit coupling on the electron count descriptors}
{We also provide some discussion on the effect of  spin-orbit coupling (SOC) on hh compounds NiMnBi and  CuNi$_{3}$Mn$_{4}$Bi$_{4}$ to show the efficacy of  our predictions. The inclusion of SOC and comparing those with spin-polarized (collinear) calculations suggests that magnetic-moment of both the hh compounds remains quenched  \cite{PhysRevB.71.012413,PhysRevB.40.3616,PhysRevB.66.094421,JPCM2004} as we did not notice any significant change in cell or atomic moments. The collinear and SOC moments of NiMnBi and CuNi$_{3}$Mn$_{4}$Bi$_{4}$ are (16.09, 16.09) $\mu_{B}$ and (17.00, 17.00) $\mu_{B}$, respectively. Similar reports on Bi based hh compounds were found \cite{GUO2016128}, where a non-vanishing DOS may appear in the minority gap by the effect of SOC but not major changes appear in electronic-structure that can impact the overall conclusion. Importantly, we also did not notice any significant change in electronic-structure or magnetic behavior arising from SOC, which indicate towards the robustness of our predictions of half-metallic nature in Table~\ref{t2n2} \cite{JPCM2004}.}

\subsection{\label{PDA} Partially disordered alloys}
The partial disorder in Co$_{2}$Mn(Sn$_{x}$Sb$_{1-x}$), (Co$_{x}$Ni$_{1-x}$)$_{2}$MnA and Co$_{2}$(Mn$_{1-x}$Fe$_{x}$)A (here, A=\{Sn or Sb\}) was considered for structural (lattice constant), mixing enthalpy, and electronic-structure in Fig.~\ref{Co2FeMnSn_E}. We emphasize that disorder on different sublattices can produce a similar shift of the E$_{F}$. Indeed, the expected similarity of the electronic structure was found between  Co$_{2}$Mn(Sn$_{x}$Sb$_{1-x}$) and Co$_{2}$(Mn$_{1-x}$Fe$_{x}$)Sn, which has been shown in Fig.~\ref{Co2MnSnSb_Co2FeMnSn_DOS}a. Importantly, we show the optimized lattice constant (a$_{0}$) and mixing enthalpies ($E_{mix}$) of these alloys in Fig.~\ref{Co2FeMnSn_E}b, where $E_{mix}$ was found small compared to $kT_{0}$~=~$23.55$~meV (where $k$ is the Boltzmann constant) at temperature $T_0 \! = \! 273.15\,$K ($0^{\circ}$C). Mixed alloys with a small negative mixing enthalpy ($-k T_0 \! < \! E_{mix} \! \le \! 0$) can be uniform at ambient $T \ge T_0$. In contrast, alloys with a small positive mixing enthalpy ($0 \! < \! E_{mix} \! < \! kT_{0}$)  can develop a compositional fluctuation, which lowers the enthalpy \cite{JPhysD51n2p024002y2018}; however, they do not segregate at $T > E_{mix} / k$.  We find that  (Co$_{x}$Ni$_{1-x}$)$_{2}$MnA  with A=\{Sn or Sb\} have small positive $E_{mix}$, see Fig.~\ref{Co2FeMnSn_E}.

\begin{figure}[htp]
\centering
\includegraphics[scale=0.45]{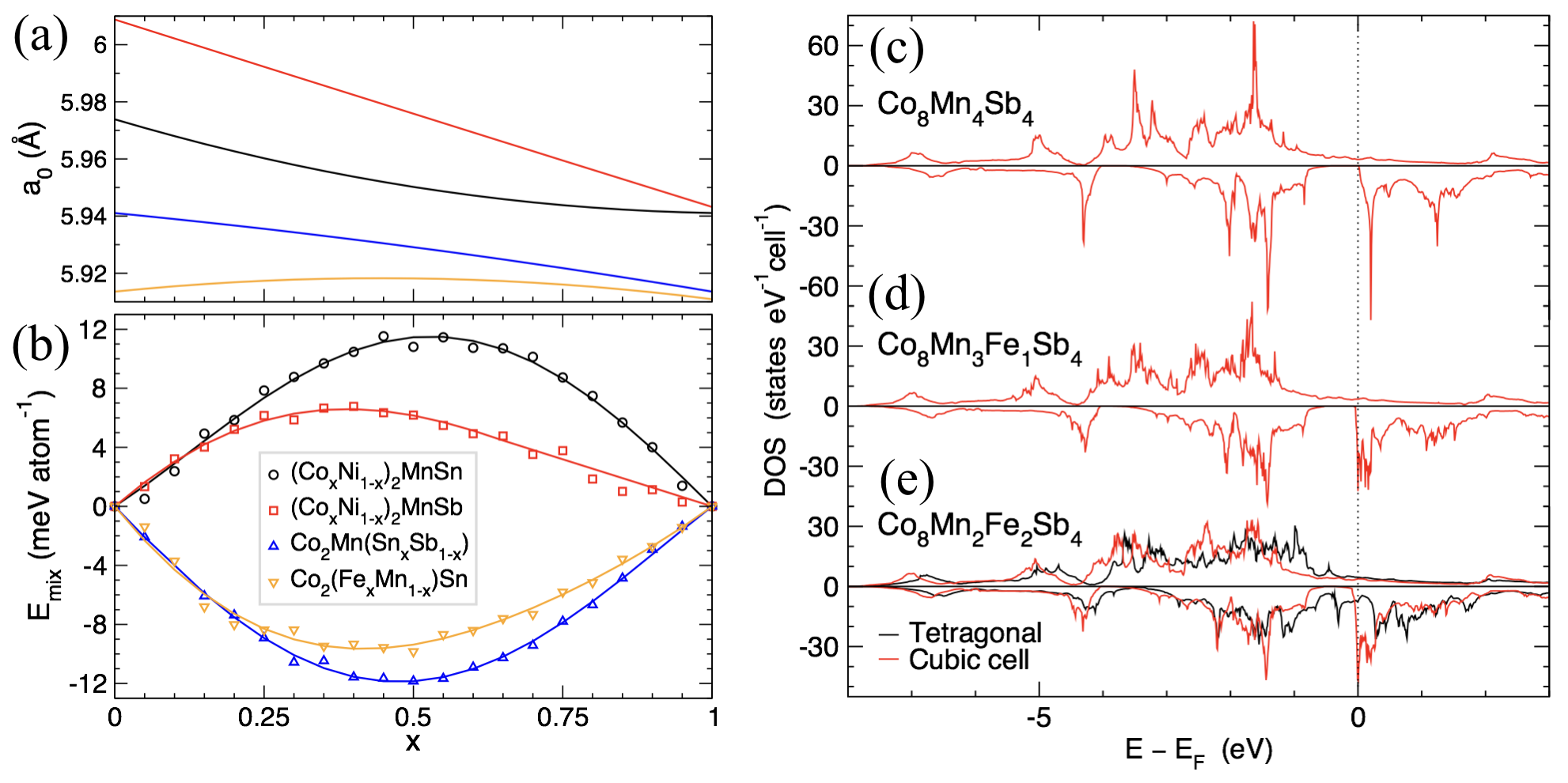} 
\caption{\label{Co2FeMnSn_E} (a,b) The equilibrium lattice constants $a_{0}$~({\angstrom}) and mixing enthalpies $E_{mix}$ (meV-atom$^{-1}$) vs. composition $x$ for fH alloys with a homogeneous chemical disorder on one of the sublattices, obtained using KKR-CPA. Lines are the piece-wise low-degree polynomial fits to DFT data (21 points, symbols). (c-e) Total spin DOS (states/eV-f.u.) of Co$_8$(Mn$_{4-4x}$Fe$_{4x}$)Sb$_4$  at $0 \le x \le 0.5$  for cubic Heusler (red) and tetragonal (black) 16-atom cell.} 
\end{figure}

\noindent
\textbf{Co$_{2}$Mn(Sn$_{1-x}$Sb$_{x}$)}: The calculated magnetization of Co$_{2}$Mn(Sn$_{1-x}$Sb$_{x}$) increases  with $x$ approximately linearly  from $1.258\,\mu_{B}$/atom in Co$_{2}$MnSn  to $1.500\,\mu_{B}$/atom in half-metallic Co$_{2}$MnSb.  The equilibrium lattice constants are slightly smaller than those from the Hume-Rothery rule: this indicates the mixing tendency. Indeed, the trend in mixing enthalpies, calculated for the partially-ordered (KKR-CPA) and fully-ordered (VASP and KKR) systems,  is the same, see Fig.~\ref{fig1str}c. The predicted stability of  disordered Co$_{2}$Mn(Sn$_{x}$Sb$_{1-x}$) alloys at RT agrees with the experiment \cite{JAP105n1p013716y2009}. The proposed compositional range of half-metals (given by $n$ in Table~\ref{t2n}) is confirmed by direct DFT calculations in Fig.~\ref{Co2MnSnSb_Co2FeMnSn_DOS}a, which shows that  disorder on different sublattices result into same change in $n$ that  can produce similar effects. Comparing calculations to experiment, we need to take into account that experimental samples are not always precisely stoichiometric.  In particular,  according to the RBS measurements \cite{JAP105n1p013716y2009}, the Co$_{2}$MnSn sample had a an excess Co, i.e., Co$_{2.03}$Mn$_{1.00}$Sn$_{0.97}$, while Co$_{2}$MnSb sample was Co-deficient, i.e., Co$_{1.98}$Mn$_{1.01}$Sb$_{1.01}$ (Co$_{1.94}$Mn$_{1.02}$Sb$_{1.04}$ according to ICP).  Consequently, we expect a slight difference between the equilibrium lattice constants $a$ calculated for the ideal stoichiometric crystals at $0\,$K, and measured on the off-stoichiometric samples at finite $T$. Notably, the lattice constants from calculations and experiments \cite{Buschow1983,JAP105n1p013716y2009} agree within 1\%, see Table~\ref{t2c}.

\noindent
\textbf{(Co$_{2-x}$Ni$_{x}$)MnSb}: With increasing Ni concentration, magnetization is decreasing and the lattice constant is increasing (Fig.~\ref{Co2FeMnSn_E}a). Because replacement of Co by Ni in Co$_{2}$MnSb moves $E_{F}$ away from the band gap, (Co$_{2-x}$Ni$_{x}$)MnSb are not half-metals at $x>0.03$, see Table~\ref{t2n}.

\noindent
\textbf{(Co$_{2-x}$Ni$_{x}$)MnSn}: These alloys can be half-metallic at $0.1 \! < \! x \! < \! 0.45$, see Table~\ref{t2n}. The rapidly quenched by melt-spinning Ni$_{47.3}$Mn$_{28.2}$Sn$_{24.5}$ sample was confirmed to be half-metallic \cite{JMMM386p98p2015}:  this sample has excessive 2.7 at.\%Mn replacing Ni and 0.5 at.\%Mn replacing Sn. Fig.~\ref{Co2MnSnSb_Co2FeMnSn_DOS}b  shows that doping not only shifts  $E_{F}$, but also narrows the band gap.  Due to the peak of electronic DOS at the Fermi level (see Fig.~\ref{Co2MnSnSb_Co2FeMnSn_DOS}b), cubic CoNiMnSn and Co$_3$NiMn$_2$Sn$_2$ structures  are unstable. We find that they can be stabilized by a tetragonal distortion, and our findings agree with recent calculations \cite{JPhysD49n39p395001y2016}. 

\noindent
\textbf{Co$_{2}$(Mn$_{1-x}$Fe$_{x}$)Sn}: In Co$_{2}$MnSn, the band gap in minority spins is located slightly above the Fermi level. Shifting $E_{F}$ to higher energies should change this alloy into a half-metal, see Fig.~\ref{Co2MnSnSb_Co2FeMnSn_DOS}a. Magnetization increases with $x$ from $1.26\,\mu_{B}$/atom in Co$_{2}$MnSn to $1.41\,\mu_{B}$/atom in Co$_{2}$FeSn and so does the lattice constant (although the atomic radius and magnetic moment of Mn is larger than that of Fe).

\noindent
\textbf{Co$_{2}$(Mn$_{1-x}$Fe$_{x}$)Sb}: In Co$_{2}$MnSb the $E_{F}$ is located at the edge of the band gap in minority spins (Fig.~\ref{fig5dos}). Replacement of Mn by Fe lowers magnetization from $1.50\,\mu_{B}$ in Co$_{2}$MnSb to $1.32\,\mu_{B}$/atom in Co$_{2}$FeSb, while distortion of Co$_{2}$(Mn$_{0.5}$Fe$_{0.5}$)Sb from unstable cubic to a more stable tetragonal structure lowers magnetization from 1.42 to $1.23\,\mu_{B}$/atom. An attempt to mix Mn with Fe on the Mn sub\-lattice quickly moves $E_{F}$ away from the band gap  [thus, Co$_2$(Mn$_{1-x}$Fe$_x$)Sb are not half-metals for $x>0.0$] into the large peak in the minority spins (Fig.~\ref{Co2FeMnSn_E}b-d), making a cubic Heusler structure unstable. Indeed, mixing enthalpies of the cubic Co$_{2}$(Mn$_{1-x}$Fe$_{x}$)Sb structures  are positive (relative to the segregated  Co$_{2}$MnSb and Co$_{2}$FeSb), with that for cubic Co$_{2}$(Mn$_{0.5}$Fe$_{0.5}$)Sb being $+2.1\,$meV/atom, while distorted tetragonal Co$_{2}$(Mn$_{0.5}$Fe$_{0.5}$)Sb has a negative formation enthalpy of $-10.3\,$meV/atom.  A similar instability towards tetragonal distortions was predicted for other  Co$_2$-based Heusler compounds under pressure \cite{JPhysD49n35p355004y2016}. We conclude that adding iron to Co$_{2}$MnSb can result in formation of other (more stable) compounds with lower magnetization, which are not Heusler.

\section{Summary}

In this work, we offer a design guide based on electron count for achieving half-metallicity in Heusler alloys, where the electron count was manipulated both by substitutional doping and disorder. The intelligent choice of alloying elements was found to tune the the total number of electrons thats shifts the Fermi-level to the band gap region, making the system half-metallic. The electronic-structure study of half-metallic compositions  was performed using direct DFT methods, and found that the mixing of different alloying elements on different sublattices can produce similar effects on the electronic-structure. We also found several stable Heusler alloys their compositions, e.g., Co$_{2}$-Mn-X and Ni-Mn-X (with X=\{Al, Si, Ga, Ge, Sn, Sb\}), with low mixing enthalpies. Based on the idea of electron count and disorder, we predicted that hH NiMnBi can be made half-metallic by substituting Ni by specific Cu and Zn compositions ((CuNi$_{3}$)Mn$_{4}$Bi$_{4}$ and  (ZnNi$_{7}$)Mn$_{8}$Bi$_{8}$), which also provides improved thermodynamic stability to hH NiMnBi. Our work provides prediction of numerous multicomponent half-metals with desired width of the band gap as marked in Tables~1\& 2. The uniqueness of our approach lies in the fact that it provides fast electron-count based assessment of half-metallicity. The design guide presented here will play a dominant role in discovering novel compositions, which will accelerate the development and synthesis of new materials. The idea of tuning energy-gap using electron-count can also be as used a feature to train machine-learning models that will allow high-throughput design of multicomponent half-metals.

\section{Acknowledgments}
This work was supported by the U.S. Department of Energy (DOE), Office of Science, Basic Energy Sciences, Materials Science and Engineering Division. Application of the electronic-structure methods to caloric materials is supported by the DOE's Advanced Manufacturing Office of the Office of Energy Efficiency and Renewable Energy through CaloriCool\textsuperscript{TM} -- the Caloric Materials Consortium, established as a part of the U.S. DOE Energy Materials Network. The research is performed at the Ames Laboratory, which is operated for the U.S. DOE by Iowa State University under contract DE-AC02-07CH11358.

\section{Data availability}
The authors declare that the data supporting the findings of this study are available within the paper and supplement. Also, the data that support the plots within this paper and other findings of this study are available from the corresponding author upon reasonable request.

\biboptions{sort&compress}

\bibliography{Heusler}

\begin{thebibliography}{100}
\expandafter\ifx\csname url\endcsname\relax
  \def\url#1{\texttt{#1}}\fi
\expandafter\ifx\csname urlprefix\endcsname\relax\def\urlprefix{URL }\fi
\expandafter\ifx\csname href\endcsname\relax
  \def\href#1#2{#2} \def\path#1{#1}\fi

\bibitem{Heusler1903}
{F. Heusler}, Verh. Dtsch. Phys. Ges. 5 (1903) 219.

\bibitem{JSSChem43p354y1982}
{E. Uhl}, The ferromagnetic and paramagnetic properties of heusler alloys
  (ni$_{1-x}$co$_{x}$)$_2$mnsb, J. Solid State Chem. 43 (1982) 354--358.

\bibitem{HeuslerAlloys2016}
C.~Felser, A.~Hirohata (Eds.), Heusler Alloys: Properties, Growth,
  Applications, Springer, 2016.

\bibitem{SciRep8n1p9147y2018}
B.~Weise, B.~Dutta, N.~Teichert, A.~H{\"u}tten, T.~Hickel, A.~Waske,
  \href{https://doi.org/10.1038/s41598-018-27428-8}{Role of disorder when
  upscaling magnetocaloric ni-co-mn-al heusler alloys from thin films to
  ribbons}, Scientific Reports 8~(1) (2018) 9147.
\newblock \href {http://dx.doi.org/10.1038/s41598-018-27428-8}
  {\path{doi:10.1038/s41598-018-27428-8}}.
\newline\urlprefix\url{https://doi.org/10.1038/s41598-018-27428-8}

\bibitem{ActaMat142p49y2018}
S.~Li, T.~Nakatani, K.~Masuda, Y.~Sakuraba, X.~Xu, T.~Sasaki, H.~Tajiri,
  Y.~Miura, T.~Furubayashi, K.~Hono,
  \href{http://www.sciencedirect.com/science/article/pii/S1359645417308121}{Enhancement
  of current-perpendicular-to-plane giant magnetoresistive outputs by improving
  b2-order in polycrystalline co2(mn0.6fe0.4)ge heusler alloy films with the
  insertion of amorphous cofebta underlayer}, Acta Materialia 142 (2018) 49 --
  57.
\newblock \href
  {http://dx.doi.org/https://doi.org/10.1016/j.actamat.2017.09.046}
  {\path{doi:https://doi.org/10.1016/j.actamat.2017.09.046}}.
\newline\urlprefix\url{http://www.sciencedirect.com/science/article/pii/S1359645417308121}

\bibitem{ActaMat148p216y2018}
M.~N. Guzik, C.~Echevarria-Bonet, M.~D. Riktor, P.~A. Carvalho, A.~E. Gunnaes,
  M.~H. Sorby, B.~C. Hauback,
  \href{http://www.sciencedirect.com/science/article/pii/S1359645418300508}{Half-heusler
  phase formation and ni atom distribution in m-ni-sn (m = hf, ti, zr)
  systems}, Acta Materialia 148 (2018) 216 -- 224.
\newblock \href
  {http://dx.doi.org/https://doi.org/10.1016/j.actamat.2018.01.019}
  {\path{doi:https://doi.org/10.1016/j.actamat.2018.01.019}}.
\newline\urlprefix\url{http://www.sciencedirect.com/science/article/pii/S1359645418300508}

\bibitem{ActaMat115p308y2016}
E.~Rausch, M.~V. Castegnaro, F.~Bernardi, M.~C.~M. Alves, J.~Morais, B.~Balke,
  \href{http://www.sciencedirect.com/science/article/pii/S1359645416303925}{Short
  and long range order of half-heusler phases in (ti,zr,hf)cosb thermoelectric
  compounds}, Acta Materialia 115 (2016) 308 -- 313.
\newblock \href
  {http://dx.doi.org/https://doi.org/10.1016/j.actamat.2016.05.041}
  {\path{doi:https://doi.org/10.1016/j.actamat.2016.05.041}}.
\newline\urlprefix\url{http://www.sciencedirect.com/science/article/pii/S1359645416303925}

\bibitem{ActaMat107p1359y2016}
G.~Rogl, A.~Grytsiv, M.~Gurth, A.~Tavassoli, C.~Ebner, A.~Wunschek,
  S.~Puchegger, V.~Soprunyuk, W.~Schranz, E.~Bauer, H.~Muller, M.~Zehetbauer,
  P.~Rogl,
  \href{http://www.sciencedirect.com/science/article/pii/S1359645416300301}{Mechanical
  properties of half-heusler alloys}, Acta Materialia 107 (2016) 178 -- 195.
\newblock \href
  {http://dx.doi.org/https://doi.org/10.1016/j.actamat.2016.01.031}
  {\path{doi:https://doi.org/10.1016/j.actamat.2016.01.031}}.
\newline\urlprefix\url{http://www.sciencedirect.com/science/article/pii/S1359645416300301}

\bibitem{ActaMat104p210y2016}
M.~Gurth, G.~Rogl, V.~Romaka, A.~Grytsiv, E.~Bauer, P.~Rogl,
  \href{http://www.sciencedirect.com/science/article/pii/S1359645415300732}{Thermoelectric
  high zt half-heusler alloys $ti_{1-x-y}zr_x hf_y nisn$ (0<x<1; 0<y<1)}, Acta
  Materialia 104 (2016) 210 -- 222.
\newblock \href
  {http://dx.doi.org/https://doi.org/10.1016/j.actamat.2015.11.022}
  {\path{doi:https://doi.org/10.1016/j.actamat.2015.11.022}}.
\newline\urlprefix\url{http://www.sciencedirect.com/science/article/pii/S1359645415300732}

\bibitem{PRB94p024418y2016}
J.~E. Fischer, J.~Karel, S.~Fabbrici, P.~Adler, S.~Ouardi, G.~H. Fecher,
  F.~Albertini, C.~Felser,
  \href{https://link.aps.org/doi/10.1103/PhysRevB.94.024418}{Magnetic
  properties and curie temperatures of disordered heusler compounds:
  ${\mathrm{co}}_{1+x}{\mathrm{fe}}_{2\ensuremath{-}x}\mathrm{Si}$}, Phys. Rev.
  B 94 (2016) 024418.
\newblock \href {http://dx.doi.org/10.1103/PhysRevB.94.024418}
  {\path{doi:10.1103/PhysRevB.94.024418}}.
\newline\urlprefix\url{https://link.aps.org/doi/10.1103/PhysRevB.94.024418}

\bibitem{Buschow1983}
K.~Buschow, P.~van Engen, R.~Jongebreur,
  \href{http://www.sciencedirect.com/science/article/pii/0304885383900975}{Magneto-optical
  properties of metallic ferromagnetic materials}, Journal of Magnetism and
  Magnetic Materials 38~(1) (1983) 1 -- 22.
\newblock \href
  {http://dx.doi.org/http://dx.doi.org/10.1016/0304-8853(83)90097-5}
  {\path{doi:http://dx.doi.org/10.1016/0304-8853(83)90097-5}}.
\newline\urlprefix\url{http://www.sciencedirect.com/science/article/pii/0304885383900975}

\bibitem{PJW1971}
{P.J. Webster}, Magnetic and chemical order in heusler alloys containing cobalt
  and manganese, J. Phys. Chem. Solids 32 (1971) 1221--1231.
\newblock \href
  {http://dx.doi.org/http://dx.doi.org/10.1016/S0022-3697(71)80180-4}
  {\path{doi:http://dx.doi.org/10.1016/S0022-3697(71)80180-4}}.

\bibitem{Brown2000}
P.~J. Brown, K.~U. Neumann, P.~J. Webster, K.~R.~A. Ziebeck, The magnetization
  distributions in some heusler alloys proposed as half-metallic ferromagnets,
  Journal of Physics: Condensed Matter 12 (2000) 1827--1835.
\newblock \href
  {http://dx.doi.org/http://stacks.iop.org/0953-8984/12/i=8/a=325}
  {\path{doi:http://stacks.iop.org/0953-8984/12/i=8/a=325}}.

\bibitem{JPhysD51n2p024002y2018}
N.~A. Zarkevich, D.~D. Johnson, V.~K. Pecharsky,
  \href{http://stacks.iop.org/0022-3727/51/i=2/a=024002}{High-throughput search
  for caloric materials: the caloricool approach}, Journal of Physics D:
  Applied Physics 51~(2) (2018) 024002.
\newblock \href {http://dx.doi.org/10.1088/1361-6463/aa9bd0}
  {\path{doi:10.1088/1361-6463/aa9bd0}}.
\newline\urlprefix\url{http://stacks.iop.org/0022-3727/51/i=2/a=024002}

\bibitem{JPhysD49n39p395001y2016}
A.~Gr\"unebohm, H.~C. Herper, P.~Entel,
  \href{http://stacks.iop.org/0022-3727/49/i=39/a=395001}{On the rich magnetic
  phase diagram of (ni,co)-mn-sn heusler alloys}, Journal of Physics D: Applied
  Physics 49~(39) (2016) 395001.
\newline\urlprefix\url{http://stacks.iop.org/0022-3727/49/i=39/a=395001}

\bibitem{Groot1983}
R.~A. de~Groot, F.~M. Mueller, P.~G.~v. Engen, K.~H.~J. Buschow, New class of
  materials: Half-metallic ferromagnets, Phys. Rev. Lett. 50 (1983) 2024--2027.
\newblock \href {http://dx.doi.org/10.1103/PhysRevLett.50.2024}
  {\path{doi:10.1103/PhysRevLett.50.2024}}.

\bibitem{Hanssen1986}
K.~E. H.~M. Hanssen, P.~E. Mijnarends, Positron-annihilation study of the
  half-metallic ferromagnet nimnsb: Theory, Phys. Rev. B 34 (1986) 5009--5016.
\newblock \href {http://dx.doi.org/10.1103/PhysRevB.34.5009}
  {\path{doi:10.1103/PhysRevB.34.5009}}.

\bibitem{JPhysCM1p2341y1989}
M.~J. Otto, R.~A.~M. van Woerden, P.~J. van~der Valk, J.~Wijngaard, C.~F. van
  Bruggen, C.~Haas, K.~H.~J. Buschow,
  \href{http://stacks.iop.org/0953-8984/1/i=13/a=007}{Half-metallic
  ferromagnets. i. structure and magnetic properties of nimnsb and related
  inter-metallic compounds}, Journal of Physics: Condensed Matter 1~(13) (1989)
  2341.
\newline\urlprefix\url{http://stacks.iop.org/0953-8984/1/i=13/a=007}

\bibitem{Fujii1989}
S.~Fujii, S.~Ishida, S.~Asano, Electronic structure and lattice transformation
  in ni2mnga and co2nbsn, Journal of the Physical Society of Japan 58 (1989)
  3657--3665.
\newblock \href {http://dx.doi.org/10.1143/JPSJ.58.3657}
  {\path{doi:10.1143/JPSJ.58.3657}}.

\bibitem{Fujii1990}
S.~Fujii, S.~Sugimura, Ishida, S.~Asano, Hyperfine fields and electronic
  structures of the heusler alloys co 2 mnx (x=al, ga, si, ge, sn), Journal of
  Physics: Condensed Matter 2 (1990) 8583.
\newblock \href {http://dx.doi.org/https://doi.org/10.1088/0953-8984/2/43/004}
  {\path{doi:https://doi.org/10.1088/0953-8984/2/43/004}}.

\bibitem{Fuji1995}
S.~Fujii, S.~Ishida, S.~Asano, A half-metallic band structure and fe$_{2}$mnz
  (z=al, si, p), Journal of the Physical Society of Japan 64 (1995) 185--191.
\newblock \href {http://dx.doi.org/10.1143/JPSJ.64.185}
  {\path{doi:10.1143/JPSJ.64.185}}.

\bibitem{Galanakis2002}
I.~Galanakis, P.~H. Dederichs, N.~Papanikolaou, Origin and properties of the
  gap in the half-ferromagnetic heusler alloys, Phys. Rev. B 66 (2002) 134428.
\newblock \href {http://dx.doi.org/10.1103/PhysRevB.66.134428}
  {\path{doi:10.1103/PhysRevB.66.134428}}.

\bibitem{Galanakis2002_1}
{Galanakis, I. and Dederichs, P. H. and Papanikolaou, N.}, Slater-pauling
  behavior and origin of the half-metallicity of the full-heusler alloys, Phys.
  Rev. B 66 (2002) 174429.
\newblock \href {http://dx.doi.org/10.1103/PhysRevB.66.174429}
  {\path{doi:10.1103/PhysRevB.66.174429}}.

\bibitem{NatureComm5p3682y2014}
{G.-X. Miao, J. Chang, B.A. Assaf, D. Heiman, and J. S. Moodera}, Spin
  regulation in composite spin-filter barrier devices, Nature Comm. 5 (2014)
  3682.
\newblock \href {http://dx.doi.org/10.1038/ncomms4682}
  {\path{doi:10.1038/ncomms4682}}.

\bibitem{Krishnan2016}
K.~M. Krishnan,
  \href{http://www.oxfordscholarship.com/view/10.1093/acprof:oso/9780199570447.001.0001/acprof-9780199570447}{Fundamentals
  and Applications of Magnetic Materials}, Oxford University Press, 2016.
\newblock \href {http://dx.doi.org/10.1093/acprof:oso/9780199570447.001.0001}
  {\path{doi:10.1093/acprof:oso/9780199570447.001.0001}}.
\newline\urlprefix\url{http://www.oxfordscholarship.com/view/10.1093/acprof:oso/9780199570447.001.0001/acprof-9780199570447}

\bibitem{Science235n4785p172y1987}
F.~CAPASSO, \href{http://science.sciencemag.org/content/235/4785/172}{Band-gap
  engineering: From physics and materials to new semiconductor devices},
  Science 235~(4785) (1987) 172--176.
\newblock \href
  {http://arxiv.org/abs/http://science.sciencemag.org/content/235/4785/172.full.pdf}
  {\path{arXiv:http://science.sciencemag.org/content/235/4785/172.full.pdf}},
  \href {http://dx.doi.org/10.1126/science.235.4785.172}
  {\path{doi:10.1126/science.235.4785.172}}.
\newline\urlprefix\url{http://science.sciencemag.org/content/235/4785/172}

\bibitem{Science352n6292p1446y2016}
M.~Schwarze, W.~Tress, B.~Beyer, F.~Gao, R.~Scholz, C.~Poelking, K.~Ortstein,
  A.~A. G{\"u}nther, D.~Kasemann, D.~Andrienko, K.~Leo,
  \href{http://science.sciencemag.org/content/352/6292/1446}{Band structure
  engineering in organic semiconductors}, Science 352~(6292) (2016) 1446--1449.
\newblock \href
  {http://arxiv.org/abs/http://science.sciencemag.org/content/352/6292/1446.full.pdf}
  {\path{arXiv:http://science.sciencemag.org/content/352/6292/1446.full.pdf}},
  \href {http://dx.doi.org/10.1126/science.aaf0590}
  {\path{doi:10.1126/science.aaf0590}}.
\newline\urlprefix\url{http://science.sciencemag.org/content/352/6292/1446}

\bibitem{Science294p1488y2001}
{S. A. Wolf, D. D. Awschalom, R. A. Buhrman, J. M. Daughton, S. von Moln\'ar,
  M. L. Roukes, A. Y. Chtchelkanova, D. M. Treger}, Spintronics: A spin-based
  electronics vision for the future, Science 294 (2001) 1488.

\bibitem{Spintronics2013}
{C. Felser and G.~H. Fecher} (Ed.), Spintronics: From Materials to Devices,
  Springer, New York, 2013.
\newblock \href {http://dx.doi.org/10.1007/978-90-481-3832-6}
  {\path{doi:10.1007/978-90-481-3832-6}}.

\bibitem{NRM3n5p5y2018}
D.~P. Tabor, L.~M. Roch, S.~K. Saikin, C.~Kreisbeck, D.~Sheberla, J.~H.
  Montoya, S.~Dwaraknath, M.~Aykol, C.~Ortiz, H.~Tribukait, C.~Amador-Bedolla,
  C.~J. Brabec, B.~Maruyama, K.~A. Persson, A.~Aspuru-Guzik,
  \href{https://doi.org/10.1038/s41578-018-0005-z}{Accelerating the discovery
  of materials for clean energy in the era of smart automation}, Nature Reviews
  Materials 3~(5) (2018) 5--20.
\newblock \href {http://dx.doi.org/10.1038/s41578-018-0005-z}
  {\path{doi:10.1038/s41578-018-0005-z}}.
\newline\urlprefix\url{https://doi.org/10.1038/s41578-018-0005-z}

\bibitem{NRM2p17053y2017}
P.~Gorai, V.~Stevanovi{\'c}, E.~S. Toberer,
  \href{https://doi.org/10.1038/natrevmats.2017.53}{Computationally guided
  discovery of thermoelectric materials}, Nature Reviews Materials 2 (2017)
  17053.
\newblock \href {http://dx.doi.org/10.1038/natrevmats.2017.53}
  {\path{doi:10.1038/natrevmats.2017.53}}.
\newline\urlprefix\url{https://doi.org/10.1038/natrevmats.2017.53}

\bibitem{NRM1p15004y2016}
A.~Jain, Y.~Shin, K.~A. Persson,
  \href{https://doi.org/10.1038/natrevmats.2015.4}{Computational predictions of
  energy materials using density functional theory}, Nature Reviews Materials 1
  (2016) 15004.
\newblock \href {http://dx.doi.org/10.1038/natrevmats.2015.4}
  {\path{doi:10.1038/natrevmats.2015.4}}.
\newline\urlprefix\url{https://doi.org/10.1038/natrevmats.2015.4}

\bibitem{JMS47n21p7317y2012}
G.~Hautier, A.~Jain, S.~P. Ong,
  \href{https://doi.org/10.1007/s10853-012-6424-0}{From the computer to the
  laboratory: materials discovery and design using first-principles
  calculations}, Journal of Materials Science 47~(21) (2012) 7317--7340.
\newblock \href {http://dx.doi.org/10.1007/s10853-012-6424-0}
  {\path{doi:10.1007/s10853-012-6424-0}}.
\newline\urlprefix\url{https://doi.org/10.1007/s10853-012-6424-0}

\bibitem{Hanssen1990}
K.~E. H.~M. Hanssen, P.~E. Mijnarends, L.~P. L.~M. Rabou, K.~H.~J. Buschow,
  Positron-annihilation study of the half-metallic ferromagnet nimnsb:
  Experiment, Phys. Rev. B 42 (1990) 1533--1540.
\newblock \href {http://dx.doi.org/10.1103/PhysRevB.42.1533}
  {\path{doi:10.1103/PhysRevB.42.1533}}.

\bibitem{Raphael2002}
M.~P. Raphael, B.~Ravel, Q.~Huang, M.~A. Willard, S.~F. Cheng, B.~N. Das, R.~M.
  Stroud, K.~M. Bussmann, J.~H. Claassen, V.~G. Harris, Presence of antisite
  disorder and its characterization in the predicted half-metal
  ${\mathrm{co}}_{2}\mathrm{MnSi}$, Phys. Rev. B 66 (2002) 104429.
\newblock \href {http://dx.doi.org/10.1103/PhysRevB.66.104429}
  {\path{doi:10.1103/PhysRevB.66.104429}}.

\bibitem{Complexity11p36y2006}
N.~A. Zarkevich, Structural database for reducing cost in materials design and
  complexity of multiscale computations, Complexity 11 (2006) 36--42.
\newblock \href {http://dx.doi.org/10.1002/cplx.20117}
  {\path{doi:10.1002/cplx.20117}}.

\bibitem{VASP1}
G.~Kresse, J.~Hafner,
  \href{https://link.aps.org/doi/10.1103/PhysRevB.47.558}{Ab initio molecular
  dynamics for liquid metals}, Phys. Rev. B 47 (1993) 558--561.
\newblock \href {http://dx.doi.org/10.1103/PhysRevB.47.558}
  {\path{doi:10.1103/PhysRevB.47.558}}.
\newline\urlprefix\url{https://link.aps.org/doi/10.1103/PhysRevB.47.558}

\bibitem{VASP2}
G.~Kresse, J.~Hafner,
  \href{https://link.aps.org/doi/10.1103/PhysRevB.49.14251}{Ab initio
  molecular-dynamics simulation of the liquid-metal--amorphous-semiconductor
  transition in germanium}, Phys. Rev. B 49 (1994) 14251--14269.
\newblock \href {http://dx.doi.org/10.1103/PhysRevB.49.14251}
  {\path{doi:10.1103/PhysRevB.49.14251}}.
\newline\urlprefix\url{https://link.aps.org/doi/10.1103/PhysRevB.49.14251}

\bibitem{PAW}
P.~E. Bl\"ochl,
  \href{https://link.aps.org/doi/10.1103/PhysRevB.50.17953}{Projector
  augmented-wave method}, Phys. Rev. B 50 (1994) 17953--17979.
\newblock \href {http://dx.doi.org/10.1103/PhysRevB.50.17953}
  {\path{doi:10.1103/PhysRevB.50.17953}}.
\newline\urlprefix\url{https://link.aps.org/doi/10.1103/PhysRevB.50.17953}

\bibitem{PAW2}
G.~Kresse, D.~Joubert,
  \href{https://link.aps.org/doi/10.1103/PhysRevB.59.1758}{From ultrasoft
  pseudopotentials to the projector augmented-wave method}, Phys. Rev. B 59
  (1999) 1758--1775.
\newblock \href {http://dx.doi.org/10.1103/PhysRevB.59.1758}
  {\path{doi:10.1103/PhysRevB.59.1758}}.
\newline\urlprefix\url{https://link.aps.org/doi/10.1103/PhysRevB.59.1758}

\bibitem{PBESol2008}
J.~P. Perdew, A.~Ruzsinszky, G.~I. Csonka, O.~A. Vydrov, G.~E. Scuseria, L.~A.
  Constantin, X.~Zhou, K.~Burke,
  \href{http://link.aps.org/doi/10.1103/PhysRevLett.100.136406}{Restoring the
  density-gradient expansion for exchange in solids and surfaces}, Phys. Rev.
  Lett. 100 (2008) 136406.
\newblock \href {http://dx.doi.org/10.1103/PhysRevLett.100.136406}
  {\path{doi:10.1103/PhysRevLett.100.136406}}.
\newline\urlprefix\url{http://link.aps.org/doi/10.1103/PhysRevLett.100.136406}

\bibitem{MonkhorstPack1976}
H.~J. Monkhorst, J.~D. Pack,
  \href{https://link.aps.org/doi/10.1103/PhysRevB.13.5188}{Special points for
  brillouin-zone integrations}, Phys. Rev. B 13 (1976) 5188--5192.
\newblock \href {http://dx.doi.org/10.1103/PhysRevB.13.5188}
  {\path{doi:10.1103/PhysRevB.13.5188}}.
\newline\urlprefix\url{https://link.aps.org/doi/10.1103/PhysRevB.13.5188}

\bibitem{PRB62p6158}
P.~E. Bl\"ochl,
  \href{https://link.aps.org/doi/10.1103/PhysRevB.62.6158}{First-principles
  calculations of defects in oxygen-deficient silica exposed to hydrogen},
  Phys. Rev. B 62 (2000) 6158--6179.
\newblock \href {http://dx.doi.org/10.1103/PhysRevB.62.6158}
  {\path{doi:10.1103/PhysRevB.62.6158}}.
\newline\urlprefix\url{https://link.aps.org/doi/10.1103/PhysRevB.62.6158}

\bibitem{KKR1}
{J. Korringa}, On the calculation of the energy of a bloch wave in a metal,
  Physica 13~(6--7) (1947) 392--400.
\newblock \href {http://dx.doi.org/10.1016/0031-8914(47)90013-X}
  {\path{doi:10.1016/0031-8914(47)90013-X}}.

\bibitem{KKR2}
W.~Kohn, N.~Rostoker,
  \href{http://link.aps.org/doi/10.1103/PhysRev.94.1111}{Solution of the
  schr\"odinger equation in periodic lattices with an application to metallic
  lithium}, Phys. Rev. 94 (1954) 1111--1120.
\newblock \href {http://dx.doi.org/10.1103/PhysRev.94.1111}
  {\path{doi:10.1103/PhysRev.94.1111}}.
\newline\urlprefix\url{http://link.aps.org/doi/10.1103/PhysRev.94.1111}

\bibitem{MECCA}
{D. D. Johnson, A. V. Smirnov, and S. N. Khan}, MECCA: Multiple-scattering
  Electronic-structure Calculations for Complex Alloys. KKR-CPA Program, ver.
  2.0, Iowa State University and Ames Laboratory, Ames, 2015.

\bibitem{AJ2009}
A.~Alam, D.~D. Johnson,
  \href{http://link.aps.org/doi/10.1103/PhysRevB.80.125123}{Optimal
  site-centered electronic structure basis set from a displaced-center
  expansion: Improved results via \textit{a priori} estimates of saddle points
  in the density}, Phys. Rev. B 80 (2009) 125123.
\newblock \href {http://dx.doi.org/10.1103/PhysRevB.80.125123}
  {\path{doi:10.1103/PhysRevB.80.125123}}.
\newline\urlprefix\url{http://link.aps.org/doi/10.1103/PhysRevB.80.125123}

\bibitem{TotEnCorr_Christensen1985}
N.~E. Christensen, S.~Satpathy,
  \href{http://link.aps.org/doi/10.1103/PhysRevLett.55.600}{Pressure-induced
  cubic to tetragonal transition in csi}, Phys. Rev. Lett. 55 (1985) 600--603.
\newblock \href {http://dx.doi.org/10.1103/PhysRevLett.55.600}
  {\path{doi:10.1103/PhysRevLett.55.600}}.
\newline\urlprefix\url{http://link.aps.org/doi/10.1103/PhysRevLett.55.600}

\bibitem{AJ2012}
A.~Alam, D.~D. Johnson,
  \href{http://link.aps.org/doi/10.1103/PhysRevB.85.144202}{Structural
  properties and relative stability of (meta)stable ordered, partially ordered,
  and disordered al-li alloy phases}, Phys. Rev. B 85 (2012) 144202.
\newblock \href {http://dx.doi.org/10.1103/PhysRevB.85.144202}
  {\path{doi:10.1103/PhysRevB.85.144202}}.
\newline\urlprefix\url{http://link.aps.org/doi/10.1103/PhysRevB.85.144202}

\bibitem{PRB90p205102y2014}
A.~Alam, S.~N. Khan, A.~V. Smirnov, D.~M. Nicholson, D.~D. Johnson,
  \href{http://link.aps.org/doi/10.1103/PhysRevB.90.205102}{Green's function
  multiple-scattering theory with a truncated basis set: An augmented-kkr
  formalism}, Phys. Rev. B 90 (2014) 205102.
\newblock \href {http://dx.doi.org/10.1103/PhysRevB.90.205102}
  {\path{doi:10.1103/PhysRevB.90.205102}}.
\newline\urlprefix\url{http://link.aps.org/doi/10.1103/PhysRevB.90.205102}

\bibitem{Broyden1988}
D.~D. Johnson,
  \href{http://link.aps.org/doi/10.1103/PhysRevB.38.12807}{Modified broyden's
  method for accelerating convergence in self-consistent calculations}, Phys.
  Rev. B 38 (1988) 12807--12813.
\newblock \href {http://dx.doi.org/10.1103/PhysRevB.38.12807}
  {\path{doi:10.1103/PhysRevB.38.12807}}.
\newline\urlprefix\url{http://link.aps.org/doi/10.1103/PhysRevB.38.12807}

\bibitem{JohnsonCPA}
D.~D. Johnson, D.~M. Nicholson, F.~J. Pinski, B.~L. Gyorffy, G.~M. Stocks,
  \href{http://link.aps.org/doi/10.1103/PhysRevLett.56.2088}{Density-functional
  theory for random alloys: Total energy within the coherent-potential
  approximation}, Phys. Rev. Lett. 56 (1986) 2088--2091.
\newblock \href {http://dx.doi.org/10.1103/PhysRevLett.56.2088}
  {\path{doi:10.1103/PhysRevLett.56.2088}}.
\newline\urlprefix\url{http://link.aps.org/doi/10.1103/PhysRevLett.56.2088}

\bibitem{JP1993}
D.~D. Johnson, F.~J. Pinski,
  \href{http://link.aps.org/doi/10.1103/PhysRevB.48.11553}{Inclusion of charge
  correlations in calculations of the energetics and electronic structure for
  random substitutional alloys}, Phys. Rev. B 48 (1993) 11553--11560.
\newblock \href {http://dx.doi.org/10.1103/PhysRevB.48.11553}
  {\path{doi:10.1103/PhysRevB.48.11553}}.
\newline\urlprefix\url{http://link.aps.org/doi/10.1103/PhysRevB.48.11553}

\bibitem{VLB}
R.~van Leeuwen, E.~J. Baerends,
  \href{https://link.aps.org/doi/10.1103/PhysRevA.49.2421}{Exchange-correlation
  potential with correct asymptotic behavior}, Phys. Rev. A 49 (1994)
  2421--2431.
\newblock \href {http://dx.doi.org/10.1103/PhysRevA.49.2421}
  {\path{doi:10.1103/PhysRevA.49.2421}}.
\newline\urlprefix\url{https://link.aps.org/doi/10.1103/PhysRevA.49.2421}

\bibitem{PS2013}
P.~Singh, M.~K. Harbola, B.~Sanyal, A.~Mookerjee,
  \href{https://link.aps.org/doi/10.1103/PhysRevB.87.235110}{Accurate
  determination of band gaps within density functional formalism}, Phys. Rev. B
  87 (2013) 235110.
\newblock \href {http://dx.doi.org/10.1103/PhysRevB.87.235110}
  {\path{doi:10.1103/PhysRevB.87.235110}}.
\newline\urlprefix\url{https://link.aps.org/doi/10.1103/PhysRevB.87.235110}

\bibitem{PS2015book}
{P. Singh, M. K. Harbola, and A. Mookerjee}, Calculation of bandgaps in
  nanomaterials using Harbola-Sahni and van Leeuwen-Baerends potentials,
  Woodhead Publishing, Massachusetts, 2015.

\bibitem{PRB93p085204y2016}
P.~Singh, M.~K. Harbola, M.~Hemanadhan, A.~Mookerjee, D.~D. Johnson,
  \href{http://link.aps.org/doi/10.1103/PhysRevB.93.085204}{Better band gaps
  with asymptotically corrected local exchange potentials}, Phys. Rev. B 93
  (2016) 085204.
\newblock \href {http://dx.doi.org/10.1103/PhysRevB.93.085204}
  {\path{doi:10.1103/PhysRevB.93.085204}}.
\newline\urlprefix\url{http://link.aps.org/doi/10.1103/PhysRevB.93.085204}

\bibitem{JPCM2017Singh}
P.~Singh, M.~K. Harbola, D.~D. Johnson,
  \href{http://stacks.iop.org/0953-8984/29/i=42/a=424001}{Better band gaps for
  wide-gap semiconductors from a locally corrected exchange-correlation
  potential that nearly eliminates self-interaction errors}, Journal of
  Physics: Condensed Matter 29~(42) (2017) 424001.
\newline\urlprefix\url{http://stacks.iop.org/0953-8984/29/i=42/a=424001}

\bibitem{PhysRevB.96.054203Singh}
B.~Sadhukhan, P.~Singh, A.~Nayak, S.~Datta, D.~D. Johnson, A.~Mookerjee,
  \href{https://link.aps.org/doi/10.1103/PhysRevB.96.054203}{Band-gap tuning
  and optical response of two-dimensional
  ${\mathrm{si}}_{x}{\mathrm{c}}_{1\text{\ensuremath{-}}x}$: A first-principles
  real-space study of disordered two-dimensional materials}, Phys. Rev. B 96
  (2017) 054203.
\newblock \href {http://dx.doi.org/10.1103/PhysRevB.96.054203}
  {\path{doi:10.1103/PhysRevB.96.054203}}.
\newline\urlprefix\url{https://link.aps.org/doi/10.1103/PhysRevB.96.054203}

\bibitem{DATTA2020125}
S.~Datta, P.~Singh, D.~Jana, C.~B. Chaudhuri, M.~K. Harbola, D.~D. Johnson,
  A.~Mookerjee,
  \href{https://www.sciencedirect.com/science/article/pii/S0008622320303316}{Exploring
  the role of electronic structure on photo-catalytic behavior of
  carbon-nitride polymorphs}, Carbon 168 (2020) 125--134.
\newblock \href
  {http://dx.doi.org/https://doi.org/10.1016/j.carbon.2020.04.008}
  {\path{doi:https://doi.org/10.1016/j.carbon.2020.04.008}}.
\newline\urlprefix\url{https://www.sciencedirect.com/science/article/pii/S0008622320303316}

\bibitem{vBH}
U.~van Barth, L.~Hedin,
  \href{http://iopscience.iop.org/article/10.1088/0022-3719/5/13/012/pdf}{A
  local exchange-correlation potential for the spin polarized case: I}, J.
  Phys. C: Solid State Phys. 15 (1972) 1629--1642.
\newblock \href
  {http://dx.doi.org/http://dx.doi.org/10.1088/0022-3719/5/13/012}
  {\path{doi:http://dx.doi.org/10.1088/0022-3719/5/13/012}}.
\newline\urlprefix\url{http://iopscience.iop.org/article/10.1088/0022-3719/5/13/012/pdf}

\bibitem{TBLMTO}
{O. Jepsen and O. K. Andersen}, The Stuttgart TB-LMTO-ASA program, version 4.7,
  Max-Planck-Institut f\"ur Festk\"orperforschung, Stuttgart, Germany, 2000.

\bibitem{HSP}
M.~K. Harbola, V.~Sahni,
  \href{https://link.aps.org/doi/10.1103/PhysRevLett.62.489}{Quantum-mechanical
  interpretation of the exchange-correlation potential of kohn-sham
  density-functional theory}, Phys. Rev. Lett. 62 (1989) 489--492.
\newblock \href {http://dx.doi.org/10.1103/PhysRevLett.62.489}
  {\path{doi:10.1103/PhysRevLett.62.489}}.
\newline\urlprefix\url{https://link.aps.org/doi/10.1103/PhysRevLett.62.489}

\bibitem{HSP1}
{V. Sahni and M.K. Harbola},
  \href{http://dx.doi.org/10.1002/qua.560382456}{Quantum-mechanical
  interpretation of the local many-body potential of density-functional
  theory}, International Journal of Quantum Chemistry 38 (1990) 569--584.
\newblock \href {http://dx.doi.org/10.1002/qua.560382456}
  {\path{doi:10.1002/qua.560382456}}.
\newline\urlprefix\url{http://dx.doi.org/10.1002/qua.560382456}

\bibitem{HSH2014}
M.~S. M.~Hemanadhan, M.~K. Harbola,
  \href{http://stacks.iop.org/0953-4075/47/i=11/a=115005}{Testing an
  excited-state energy density functional and the associated potential with the
  ionization potential theorem}, Journal of Physics B: Atomic, Molecular and
  Optical Physics 47~(11) (2014) 115005.
\newline\urlprefix\url{http://stacks.iop.org/0953-4075/47/i=11/a=115005}

\bibitem{GGA}
{J. P. Perdew}, Generalized gradient approximation for the fermion kinetic
  energy as a functional of the density, Phys. Lett. A 165 (1992) 79.

\bibitem{NatComm5p3974y2014}
{M.~Jourdan {\it et al.}}, Direct observation of half-metallicity in the
  heusler compound co2mnsi, Nat. Commun. 5 (2014) 3974.
\newblock \href {http://dx.doi.org/10.1038/ncomms4974}
  {\path{doi:10.1038/ncomms4974}}.

\bibitem{PhysRevB.95.024411}
J.~Ma, V.~I. Hegde, K.~Munira, Y.~Xie, S.~Keshavarz, D.~T. Mildebrath,
  C.~Wolverton, A.~W. Ghosh, W.~H. Butler,
  \href{https://link.aps.org/doi/10.1103/PhysRevB.95.024411}{Computational
  investigation of half-heusler compounds for spintronics applications}, Phys.
  Rev. B 95 (2017) 024411.
\newblock \href {http://dx.doi.org/10.1103/PhysRevB.95.024411}
  {\path{doi:10.1103/PhysRevB.95.024411}}.
\newline\urlprefix\url{https://link.aps.org/doi/10.1103/PhysRevB.95.024411}

\bibitem{APLMat2p032103y2014}
{N. A. Zarkevich, L.-L. Wang, and D. D. Johnson}, Anomalous magneto-structural
  behavior of mnbi explained: A path towards an improved permanent magnet, APL
  Materials 2 (2014) 032103.
\newblock \href {http://dx.doi.org/10.1063/1.4867223}
  {\path{doi:10.1063/1.4867223}}.

\bibitem{Shchukarev1961}
{S. A. Shchukarev, M. P. Morozova, and T. A. Stolyarova}, Zh. Obshch. Khim.
  31~(6) (1961) 1773--1777, transtaled: Russian Journal of General Chemistry
  {\bf 31} (6), 1657--1660 (1961).

\bibitem{JAP105n1p013716y2009}
M.~R. Paudel, C.~S. Wolfe, H.~Patton, I.~Dubenko, N.~Ali, J.~A.
  Christodoulides, S.~Stadler,
  \href{http://scitation.aip.org/content/aip/journal/jap/105/1/10.1063/1.3054291}{Magnetic
  and transport properties of co2mnsnxsb1-x heusler alloys}, Journal of Applied
  Physics 105~(1) (2009) 013716.
\newblock \href {http://dx.doi.org/http://dx.doi.org/10.1063/1.3054291}
  {\path{doi:http://dx.doi.org/10.1063/1.3054291}}.
\newline\urlprefix\url{http://scitation.aip.org/content/aip/journal/jap/105/1/10.1063/1.3054291}

\bibitem{Singh2019}
S.~Datta, P.~Singh, C.~B. Chaudhuri, D.~Jana, M.~K. Harbola, D.~D. Johnson,
  A.~Mookerjee, \href{https://doi.org/10.1088/1361-648x/ab34ad}{Simple
  correction to bandgap problems in {IV} and {III}{\textendash}v
  semiconductors: an improved, local first-principles density functional
  theory} 31~(49) (2019) 495502.
\newblock \href {http://dx.doi.org/10.1088/1361-648x/ab34ad}
  {\path{doi:10.1088/1361-648x/ab34ad}}.
\newline\urlprefix\url{https://doi.org/10.1088/1361-648x/ab34ad}

\bibitem{PRB28p1745y1983}
J.~K\"ubler, A.~R. William, C.~B. Sommers,
  \href{http://link.aps.org/doi/10.1103/PhysRevB.28.1745}{Formation and
  coupling of magnetic moments in heusler alloys}, Phys. Rev. B 28 (1983)
  1745--1755.
\newblock \href {http://dx.doi.org/10.1103/PhysRevB.28.1745}
  {\path{doi:10.1103/PhysRevB.28.1745}}.
\newline\urlprefix\url{http://link.aps.org/doi/10.1103/PhysRevB.28.1745}

\bibitem{Webster1969}
P.~J. Webster, Heusler alloys, Contemporary Physics 10 (1969) 559--577.
\newblock \href {http://dx.doi.org/10.1080/00107516908204800}
  {\path{doi:10.1080/00107516908204800}}.

\bibitem{FLAPW2002_Picozzi}
S.~Picozzi, A.~Continenza, A.~J. Freeman,
  \href{http://link.aps.org/doi/10.1103/PhysRevB.66.094421}{${\mathrm{co}}_{2}\mathrm{Mn}x$
  $(x=\mathrm{Si},$ ge, sn) heusler compounds: An \textit{ab initio} study of
  their structural, electronic, and magnetic properties at zero and elevated
  pressure}, Phys. Rev. B 66 (2002) 094421.
\newblock \href {http://dx.doi.org/10.1103/PhysRevB.66.094421}
  {\path{doi:10.1103/PhysRevB.66.094421}}.
\newline\urlprefix\url{http://link.aps.org/doi/10.1103/PhysRevB.66.094421}

\bibitem{NatPhys12p855y2016}
C.~Ciccarelli, L.~Anderson, V.~Tshitoyan, A.~J. Ferguson, F.~Gerhard, C.~Gould,
  L.~W. Molenkamp, J.~Gayles, J.~Zelezny, L.~Smejkal, Z.~Yuan, J.~Sinova,
  F.~Freimuth, T.~Jungwirth, Room-temperature spin–orbit torque in nimnsb,
  Nature Physics 12 (2016) 855--860.
\newblock \href {http://dx.doi.org/10.1038/NPHYS3772}
  {\path{doi:10.1038/NPHYS3772}}.

\bibitem{PRB68p104430y2003}
L.~Ritchie, G.~Xiao, Y.~Ji, T.~Y. Chen, C.~L. Chien, M.~Zhang, J.~Chen, Z.~Liu,
  G.~Wu, X.~X. Zhang,
  \href{https://link.aps.org/doi/10.1103/PhysRevB.68.104430}{Magnetic,
  structural, and transport properties of the heusler alloys
  ${\mathrm{co}}_{2}\mathrm{MnSi}$ and nimnsb}, Phys. Rev. B 68 (2003) 104430.
\newblock \href {http://dx.doi.org/10.1103/PhysRevB.68.104430}
  {\path{doi:10.1103/PhysRevB.68.104430}}.
\newline\urlprefix\url{https://link.aps.org/doi/10.1103/PhysRevB.68.104430}

\bibitem{PRB81p054422y2010}
H.~Allmaier, L.~Chioncel, E.~Arrigoni, M.~I. Katsnelson, A.~I. Lichtenstein,
  \href{https://link.aps.org/doi/10.1103/PhysRevB.81.054422}{Half-metallicity
  in nimnsb: A variational cluster approach with ab initio parameters}, Phys.
  Rev. B 81 (2010) 054422.
\newblock \href {http://dx.doi.org/10.1103/PhysRevB.81.054422}
  {\path{doi:10.1103/PhysRevB.81.054422}}.
\newline\urlprefix\url{https://link.aps.org/doi/10.1103/PhysRevB.81.054422}

\bibitem{Bugayev1971}
{K. Bugayev, Y. Bychkov, Y. Konovalov, V. Kovalenko, and E. Tretyakov}, Iron
  and Steel Production, MIR, Moscow, 1971.

\bibitem{PRB91p174104}
N.~A. Zarkevich, D.~D. Johnson, Coexistence pressure for a martensitic
  transformation from theory and experiment: Revisiting the bcc-hcp transition
  of iron under pressure, Phys. Rev. B 91~(17) (2015) 174104.
\newblock \href {http://dx.doi.org/10.1103/PhysRevB.91.174104}
  {\path{doi:10.1103/PhysRevB.91.174104}}.

\bibitem{JChemPhys142p064707y2015}
N.~A. Zarkevich, D.~D. Johnson, Magneto-structural transformations via a
  solid-state nudged elastic band method: Application to iron under pressure,
  J. Chem. Phys. 143 (2015) 064707.
\newblock \href {http://dx.doi.org/10.1063/1.4927778}
  {\path{doi:10.1063/1.4927778}}.

\bibitem{Campbell1994}
P.~Campbell, Permanent magnet materials and their applications, Cambridge
  University Press, Cambridge, New York, Melbourne, 1994.

\bibitem{PRB93p020104y2016}
N.~A. Zarkevich, D.~D. Johnson, Titanium
  $\ensuremath{\alpha}\text{-}\ensuremath{\omega}$ phase transformation pathway
  and a predicted metastable structure, Phys. Rev. B 93 (2016) 020104.
\newblock \href {http://dx.doi.org/10.1103/PhysRevB.93.020104}
  {\path{doi:10.1103/PhysRevB.93.020104}}.

\bibitem{PRL113p265701y2014}
N.~A. Zarkevich, D.~D. Johnson, Shape-memory transformations of niti:
  Minimum-energy pathways between austenite, martensites, and kinetically
  limited intermediate states, Phys. Rev. Lett. 113 (2014) 265701.
\newblock \href {http://dx.doi.org/10.1103/PhysRevLett.113.265701}
  {\path{doi:10.1103/PhysRevLett.113.265701}}.

\bibitem{PRB90p060102y2014}
N.~A. Zarkevich, D.~D. Johnson, Stable atomic structure of niti austenite,
  Phys. Rev. B 90 (2014) 060102.
\newblock \href {http://dx.doi.org/10.1103/PhysRevB.90.060102}
  {\path{doi:10.1103/PhysRevB.90.060102}}.

\bibitem{C2NEB}
{N. A. Zarkevich and D. D. Johnson}, Nudged-elastic band method with two
  climbing images: Finding transition states in complex energy landscapes, J.
  Chem. Phys. 142~(2) (2015) 024106.
\newblock \href {http://dx.doi.org/http://dx.doi.org/10.1063/1.4905209}
  {\path{doi:http://dx.doi.org/10.1063/1.4905209}}.

\bibitem{JChemPhys38n3p631y1963}
{F. P. Bundy}, Direct conversion of graphite to diamond in static pressure
  apparatus, J. Chem. Phys. 38~(3) (1963) 631.
\newblock \href {http://dx.doi.org/10.1063/1.1733716}
  {\path{doi:10.1063/1.1733716}}.

\bibitem{Predel1972}
{B. Predel, H. Ruge}, Bildungsenthalpien und bindungsverhältnisse in einigen
  intermetallischen verbindungen vom nias-typ, Thermochimica Acta 3~(5) (1972)
  411--418.
\newblock \href {http://dx.doi.org/10.1016/0040-6031(72)87055-2}
  {\path{doi:10.1016/0040-6031(72)87055-2}}.

\bibitem{VILLARREAL2021437}
R.~Villarreal, P.~Singh, R.~Arroyave,
  \href{https://www.sciencedirect.com/science/article/pii/S1359645420308533}{Metric-driven
  search for structurally stable inorganic compounds}, Acta Materialia 202
  (2021) 437--447.
\newblock \href
  {http://dx.doi.org/https://doi.org/10.1016/j.actamat.2020.10.055}
  {\path{doi:https://doi.org/10.1016/j.actamat.2020.10.055}}.
\newline\urlprefix\url{https://www.sciencedirect.com/science/article/pii/S1359645420308533}

\bibitem{phonopy}
A.~Togo, I.~Tanaka, First principles phonon calculations in materials science,
  Scr. Mater. 108 (2015) 1--5.

\bibitem{PhysRevB.71.012413}
I.~Galanakis,
  \href{https://link.aps.org/doi/10.1103/PhysRevB.71.012413}{Orbital magnetism
  in the half-metallic heusler alloys}, Phys. Rev. B 71 (2005) 012413.
\newblock \href {http://dx.doi.org/10.1103/PhysRevB.71.012413}
  {\path{doi:10.1103/PhysRevB.71.012413}}.
\newline\urlprefix\url{https://link.aps.org/doi/10.1103/PhysRevB.71.012413}

\bibitem{PhysRevB.40.3616}
M.~Methfessel, A.~T. Paxton,
  \href{https://link.aps.org/doi/10.1103/PhysRevB.40.3616}{High-precision
  sampling for brillouin-zone integration in metals}, Phys. Rev. B 40 (1989)
  3616--3621.
\newblock \href {http://dx.doi.org/10.1103/PhysRevB.40.3616}
  {\path{doi:10.1103/PhysRevB.40.3616}}.
\newline\urlprefix\url{https://link.aps.org/doi/10.1103/PhysRevB.40.3616}

\bibitem{PhysRevB.66.094421}
S.~Picozzi, A.~Continenza, A.~J. Freeman,
  \href{https://link.aps.org/doi/10.1103/PhysRevB.66.094421}{${\mathrm{co}}_{2}\mathrm{Mn}x$
  $(x=\mathrm{Si},$ ge, sn) heusler compounds: An ab initio study of their
  structural, electronic, and magnetic properties at zero and elevated
  pressure}, Phys. Rev. B 66 (2002) 094421.
\newblock \href {http://dx.doi.org/10.1103/PhysRevB.66.094421}
  {\path{doi:10.1103/PhysRevB.66.094421}}.
\newline\urlprefix\url{https://link.aps.org/doi/10.1103/PhysRevB.66.094421}

\bibitem{JPCM2004}
P.~Mavropoulos, I.~Galanakis, V.~Popescu, P.~H. Dederichs,
  \href{https://doi.org/10.1088/0953-8984/16/48/043}{The influence of
  spin{\textendash}orbit coupling on the band gap of heusler alloys} 16~(48)
  (2004) S5759--S5762.
\newblock \href {http://dx.doi.org/10.1088/0953-8984/16/48/043}
  {\path{doi:10.1088/0953-8984/16/48/043}}.
\newline\urlprefix\url{https://doi.org/10.1088/0953-8984/16/48/043}

\bibitem{GUO2016128}
S.-D. Guo,
  \href{https://www.sciencedirect.com/science/article/pii/S0925838815319307}{Importance
  of spin?orbit coupling in power factor calculations for half-heusler anib
  (a=ti, hf, sc, y; bsn, sb, bi)}, Journal of Alloys and Compounds 663 (2016)
  128--133.
\newblock \href
  {http://dx.doi.org/https://doi.org/10.1016/j.jallcom.2015.12.139}
  {\path{doi:https://doi.org/10.1016/j.jallcom.2015.12.139}}.
\newline\urlprefix\url{https://www.sciencedirect.com/science/article/pii/S0925838815319307}

\bibitem{JMMM386p98p2015}
M.~Nazmunnahar, T.~Ryba, J.~del Val, M.~Ipatov, J.~Gonz\'{a}lez,
  V.~Ha\v{s}kov\'{a}, P.~Szab\'{o}, P.~Samuely, J.~Kravcak, Z.~Vargova,
  R.~Varga,
  \href{http://www.sciencedirect.com/science/article/pii/S0304885315002875}{Half-metallic
  ni2mnsn heusler alloy prepared by rapid quenching}, Journal of Magnetism and
  Magnetic Materials 386~(Supplement C) (2015) 98 -- 101.
\newblock \href {http://dx.doi.org/https://doi.org/10.1016/j.jmmm.2015.03.066}
  {\path{doi:https://doi.org/10.1016/j.jmmm.2015.03.066}}.
\newline\urlprefix\url{http://www.sciencedirect.com/science/article/pii/S0304885315002875}

\bibitem{JPhysD49n35p355004y2016}
M.~A. Zagrebin, V.~V. Sokolovskiy, V.~D. Buchelnikov,
  \href{http://stacks.iop.org/0022-3727/49/i=35/a=355004}{Electronic and
  magnetic properties of the co2-based heusler compounds under pressure:
  first-principles and monte carlo studies}, Journal of Physics D: Applied
  Physics 49~(35) (2016) 355004.
\newline\urlprefix\url{http://stacks.iop.org/0022-3727/49/i=35/a=355004}

\bibitem{PRB97p014202y2018}
N.~A. Zarkevich, D.~D. Johnson,
  \href{https://link.aps.org/doi/10.1103/PhysRevB.97.014202}{Ferh ground state
  and martensitic transformation}, Phys. Rev. B 97 (2018) 014202.
\newblock \href {http://dx.doi.org/10.1103/PhysRevB.97.014202}
  {\path{doi:10.1103/PhysRevB.97.014202}}.
\newline\urlprefix\url{https://link.aps.org/doi/10.1103/PhysRevB.97.014202}

\bibitem{JMMM65p76y1987}
T.~Kanomata, K.~Shirakawa, T.~Kaneko,
  \href{http://www.sciencedirect.com/science/article/pii/030488538790312X}{Effect
  of hydrostatic pressure on the curie temperature of the heusler alloys
  ni2mnz(z = al, ga, in, sn and sb)}, Journal of Magnetism and Magnetic
  Materials 65~(1) (1987) 76 -- 82.
\newblock \href
  {http://dx.doi.org/http://dx.doi.org/10.1016/0304-8853(87)90312-X}
  {\path{doi:http://dx.doi.org/10.1016/0304-8853(87)90312-X}}.
\newline\urlprefix\url{http://www.sciencedirect.com/science/article/pii/030488538790312X}

\bibitem{Uhl1982}
{E. Uhl}, The ferromagnetic and paramagnetic properties of heusler alloys
  (ni-co)2mnsn, Journal of Solid State Chemistry 43 (1982) 354--358.

\bibitem{TiNiSi1998}
G.~A. Landrum, R.~Hoffmann, J.~Evers, H.~Boysen, The tinisi family of
  compounds: Structure and bonding, Inorganic Chemistry 37~(22) (1998)
  5754--5763.
\newblock \href {http://dx.doi.org/10.1021/ic980223e}
  {\path{doi:10.1021/ic980223e}}.

\bibitem{Matveyeva1968}
{Matveyeva N.M., Nikitina S.V., and Zezin S.B.}, Investigation of the quasi
  binary systems mnsn2-fesn2, mnsn2-cosn2 and fesn2-cosn2, Russ. Metall. 5
  (1968) 132--134.

\end{thebibliography}
\bibliographystyle{rsc}

\pagebreak
\begin{center}
\textbf{\large Supplemental Material: Effect of substitutional doping and disorder on the phase stability, magnetism, and half-metallicity of Heusler alloys}
\end{center}
\setcounter{equation}{0}
\setcounter{figure}{0}
\setcounter{table}{0}
\setcounter{page}{1}
\makeatletter

\noindent
{\it Metallic Ni$_2$MnX full Heusler alloys:}~We claim that there are no  half-metals among the fH Ni$_2$MnX alloys. We find that the ferromagnetic Ni$_2$MnX fH alloys (with X from groups 13--16 and periods 3--6) are metallic, see Fig.~\ref{fig7dos4}. They have high magnetization (Fig.~\ref{fig2M}a); some of them are promising phases for advanced Alnico-type magnets, while several might segregate into other compounds (Fig.~\ref{fig4E2}a). The electronic-structure calculations reveal  a minimum with a small density of the minority spin states, located $\approx \! 1\,$eV below the Fermi level, see Fig.~\ref{DOS_Ni2MnX_fH}.  We found that there are no half-metals among these alloys, and we do not see a practical way of transforming them into half-metals by a small amount of doping. Thus, we disagree with the suggestion that the rapidly quenched Ni$_{47.3}$Mn$_{28.2}$Sn$_{24.5}$ ribbon, prepared by melt-spinning, was half-metallic \cite{JMMM386p98p2015}.  We point that although ferromagnets can have different conductivity for two spin channels \cite{PRB97p014202y2018}, conductivity of half-metals for one of the spins is zero.  We find that Ni$_2$MnSn is ferromagnetic, but not half-metallic. The fH~Ni$_{2}$MnX alloys with X=\{P, As, Sb, Bi; S, Se, Te\}  are unstable with respect to nickel segregation (see Fig.~\ref{fig4E2}a), because they have a positive formation energy relative to the segregated metallic fcc Ni and hH NiMnX. 

\begin{figure}[htp]
\centering
\includegraphics[scale=0.5]{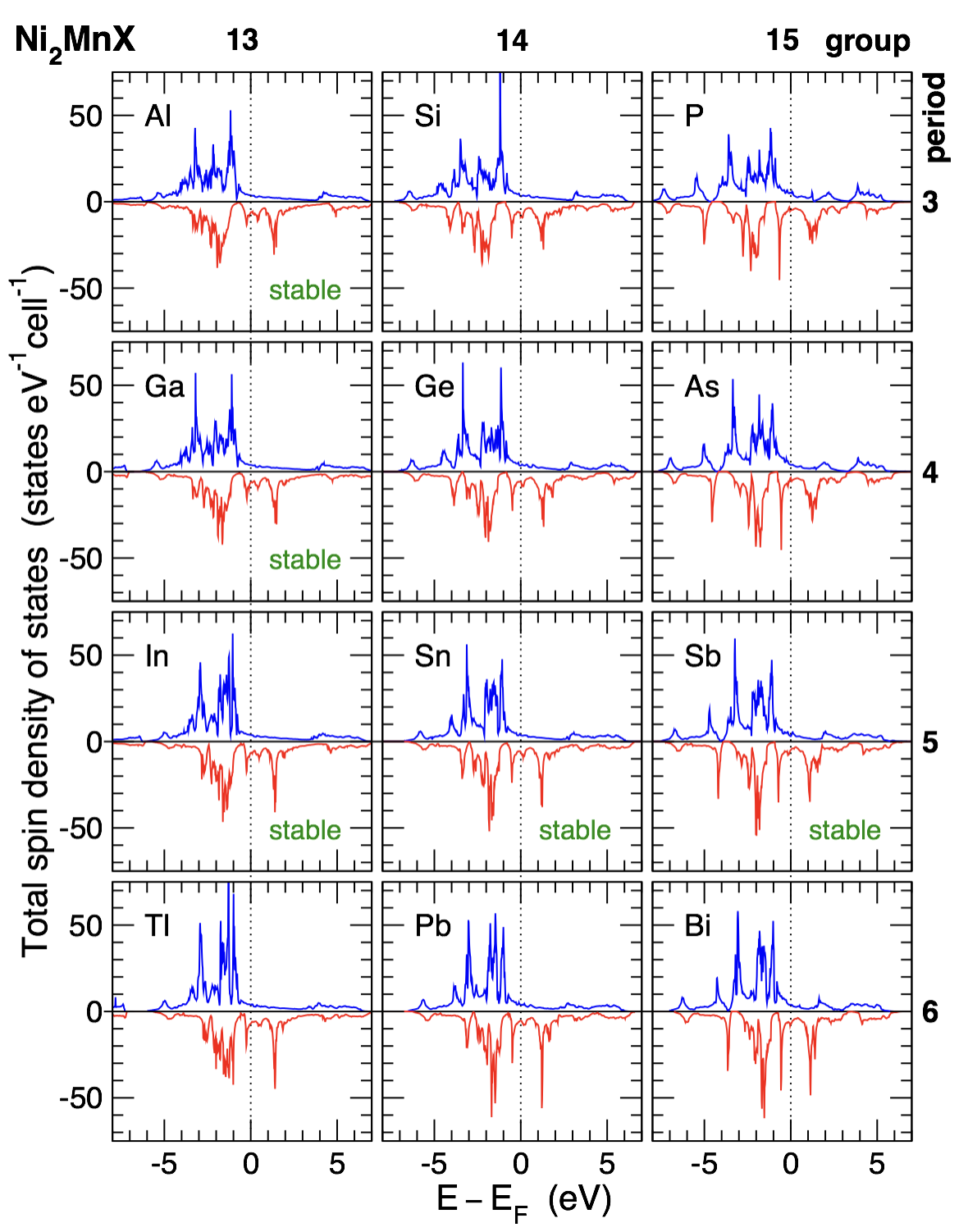} 
\caption{\label{DOS_Ni2MnX_fH} 
Total spin DOS (states/eV$\cdot$f.u.) of fH Ni$_{2}$MnX alloys for 12 elements X from group 13--15 and periods 3--6). All systems are metallic. Stable compounds are known from experiment \cite{JMMM65p76y1987,Uhl1982}.}
\end{figure}

\noindent
{\it Mechanical Distortion:~} We considered anisotropic distortions of the cubic cells, and found that  fH alloys (including Ni$_{2}$MnSb and Ni$_{2}$MnBi) might distort along the 111 axis with energy lowering,  but remain unstable with respect to segregation to fcc Ni and a hH alloy. 

\begin{figure}[htp]
\centering
\includegraphics[scale=0.4]{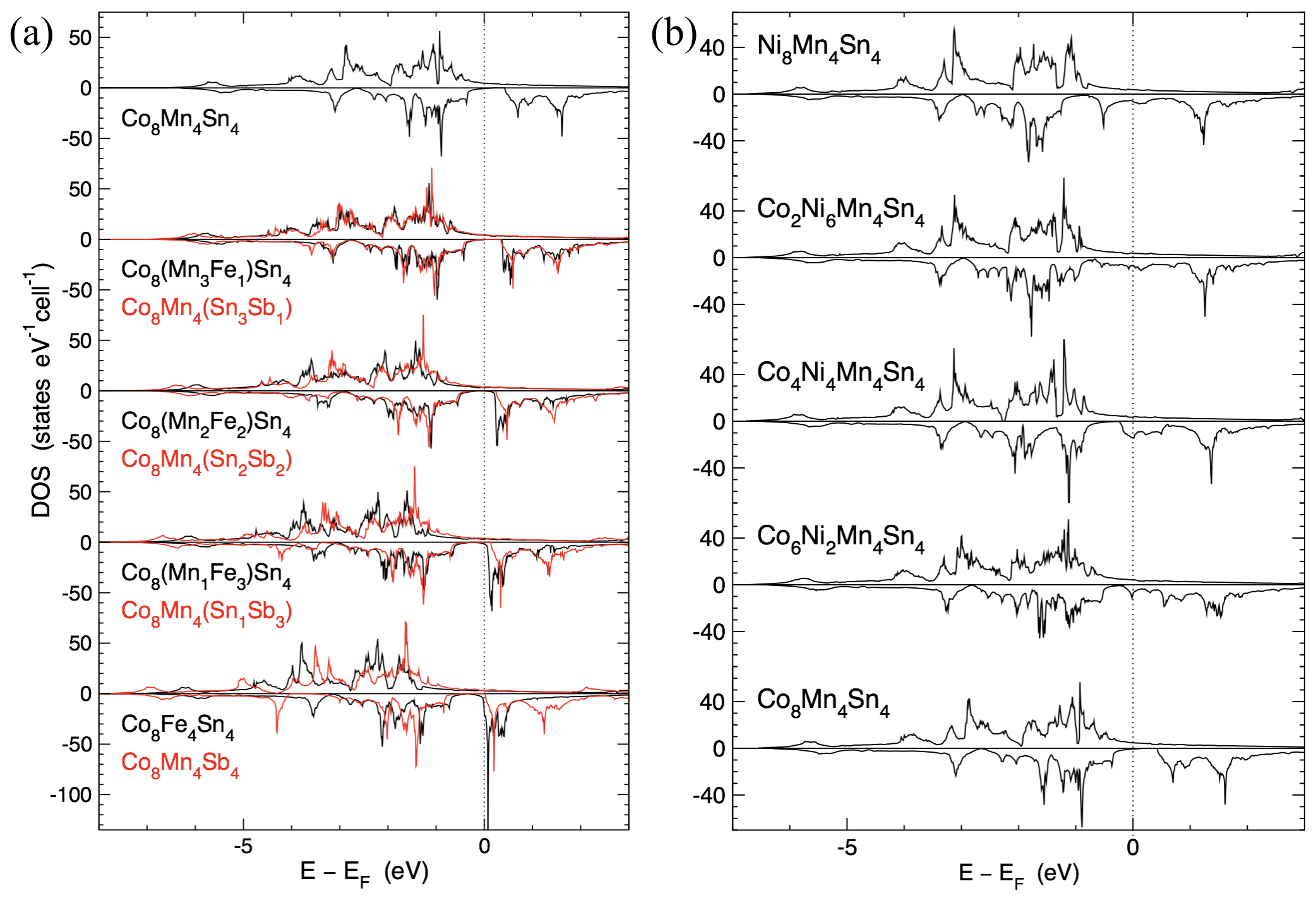}  
\caption{\label{Co2MnSnSb_Co2FeMnSn_DOS} (Left-panel) Total spin DOS of Heusler-type Co$_{8}$Mn$_{4}$(Sn$_{4x}$Sb$_{4-4x}$)
and Co$_{8}$(Mn$_{4-4x}$Fe$_{4x}$)Sn$_{4}$ fully ordered cubic structures with 16 atoms per unit cell. (Right-panel) Total spin DOS (states/eV$\cdot$f.u.) of (Co$_{2-x}$Ni$_{x}$)MnSn. The cubic Heusler structures with 16-atom unit cell.}
\end{figure}


\newpage 

\noindent
{{\it Phonon calculations of NiMnBi:} The phonon dispersion analysis of half-metallic NiMnBi was done using phonopy code \cite{phonopy}. The phonon in Fig.~\ref{Phonon_NiMnBi} shows no imaginary modes, which suggests that the predicted hH compound NiMnBi is dynamically stable, which further establishes the robustness of our prediction.} 

\begin{figure}[htp]
\centering
\includegraphics[scale=0.4]{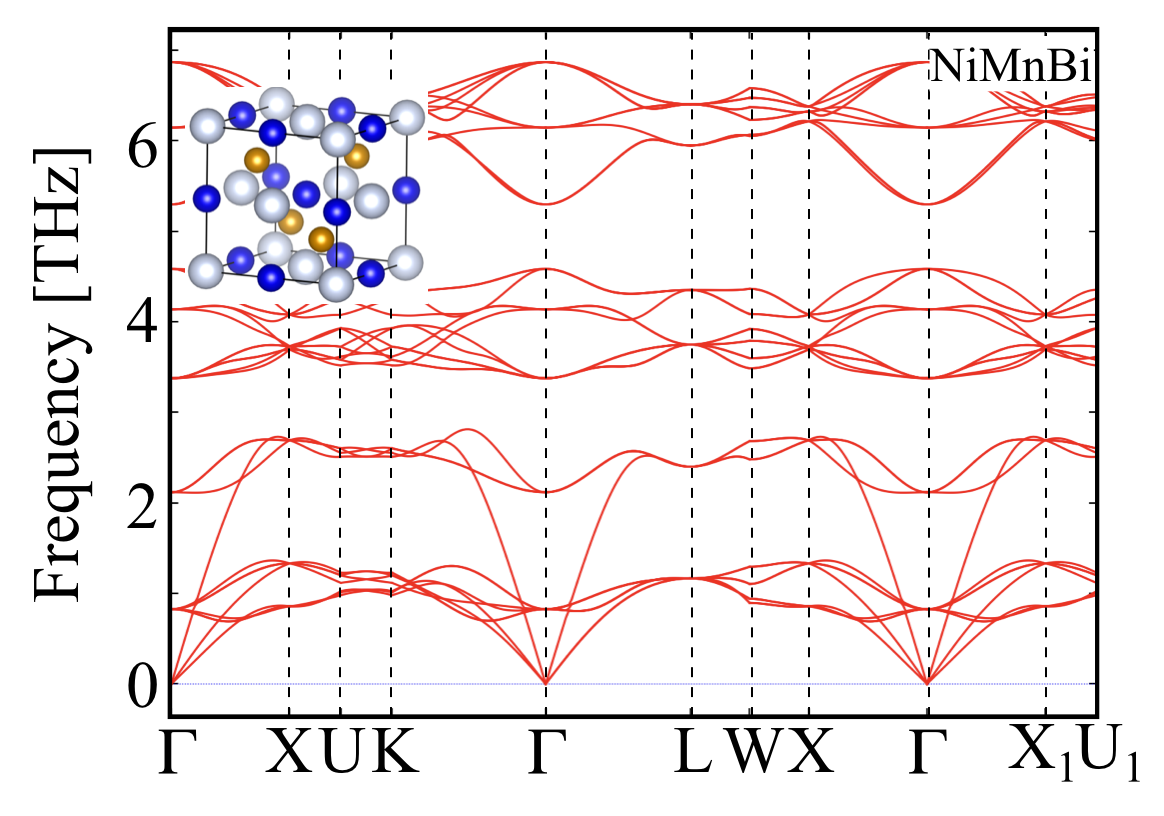}  
\caption{\label{Phonon_NiMnBi} {The phonon dispersion plot of predicted half-Heusler NiMnBi along C1b Brillouin zone ($\Gamma$-X-U-K-$\Gamma$-L-W-X-$\Gamma$-X$_{1}$-U$_1$) shows imaginary modes, i.e., the predicted structure is dynamically stable.}}
\end{figure}


\begin{table}[htp]
\centering
\caption{Equilibrium lattice constant $a$ ({\angstrom}), magnetic moment $M$ ($\mu_{B}$/f.u.), and experimental \cite{Buschow1983,PJW1971} Curie temperature $T_{c}$ (K) of fully relaxed fH Co$_{2}$MnX alloys from theory (DFT) and experiments. Asterisk ($^{*}$) marks hypothetical non-existent compounds. Known competing structures include oP12 \cite{TiNiSi1998} for X=\{Si, As\} and tI12 \cite{Matveyeva1968} for Co$_{2}$MnSn.}
\begin{tabular}{l|c|c|c|l|ll}
\hline 
\multirow{2}{*}{X } & \multicolumn{2}{c|}{\emph{a(}{\angstrom}\emph{) }} & \multicolumn{2}{c|}{ \emph{M(}$\mu_{B}$/\emph{f.u.)} } & \emph{$T_{c}$(K)}  &  \tabularnewline
\cline{2-7} 
 & DFT  & Expt. \cite{Buschow1983,PJW1971}  & DFT  & Expt.   & Expt.   &  \tabularnewline
\hline 
Al  & 5.6927  & 5.749\cite{Buschow1983}, 5.756\cite{PJW1971}  & 4.03  & 4.04, 4.01 & 693 & \tabularnewline
Ga  & 5.7136  & 5.767\cite{Buschow1983}, 5.770\cite{PJW1971}  & 4.09  & 4.05, 4.05 & 694  & \tabularnewline
In$^{*}$  & 5.9813  & \textendash{}  & 4.44  & \textendash{} & \textendash{} & \tabularnewline
Tl$^{*}$  & 6.0561  & \textendash{}  & 4.80  & \textendash{} & \textendash{} & \tabularnewline
\hline 
Si  & 5.6285  & 5.645\cite{Buschow1983}, 5.654\cite{PJW1971} & 5.00  & 4.90, 5.07 & 985 &  \tabularnewline
Ge  & 5.7358  & 5.749\cite{Buschow1983}, 5.743\cite{PJW1971} & 5.00  & 4.93, 5.11& 905  & \tabularnewline
Sn  & 5.9854  & 5.984\cite{Buschow1983}, 6.000\cite{PJW1971} & 5.03  & 4.79, 5.08 & 829  &  \tabularnewline
Pb$^{*}$  & 6.0956  & \textendash{}  & 5.11  & \textendash{} & \textendash{} & \tabularnewline
\hline 
P$^{*}$  & 5.6385  & \textendash{}   & 6.00  & \textendash{} & \textendash{} & \tabularnewline   
As$^{*}$  & 5.7939  & \textendash{}  & 5.99  & \textendash{} & \textendash{} &  \tabularnewline
Sb  & 6.0182  & 5.943\cite{JAP105n1p013716y2009}  & 6.00 & 4.52\cite{Buschow1983}  & \textendash{} & \tabularnewline
Bi$^{*}$  & 6.1793  & \textendash{}  & 6.00  & \textendash{} & \textendash{} & \tabularnewline
\hline 
\end{tabular}
\label{t2c} 
\end{table}

\begin{table}[htp]
\centering
\caption{Lattice constant $a$ ({\angstrom}) of fully relaxed hH NiMnX alloys from theory (VASP--GGA) and
experiment \cite{Buschow1983,JMMM65p76y1987,Uhl1982}. Other observed competing structures are listed in the last column.}
\begin{tabular}{|c|cc|c|}
\hline 
 \multirow{2}{*}{X } & NiMnX & hH  & Other \tabularnewline
 & GGA  & Expt.  & Expt. \tabularnewline

\hline 
Al  & 5.6078  & -  & \tabularnewline
Ga  & 5.6182  &  & \tabularnewline
In  & 5.9521  &  & \tabularnewline
Tl  & 6.0449  &  & \tabularnewline
\hline 
Si  & 5.44837  &  & oP12 \cite{TiNiSi1998} \tabularnewline
Ge & 5.570348  &  & oP12, hP6 \tabularnewline
Sn  & 5.8903  & -  & \tabularnewline
Pb  & 6.0412  &  & \tabularnewline
\hline 
P  & 5.46467  &  & oP12, hP9 \tabularnewline
As  & 5.637066  &  & oP12, hP9 \tabularnewline
Sb  & 5.9068  & 5.920 \cite{Buschow1983}  & NiMnSb$_{2}$ hP4 \tabularnewline
Bi  & 6.08546  &  & {[}cF88{]} \tabularnewline
\hline 
\end{tabular}
\label{t3a}
\end{table}

\begin{table}[htp]
\centering
\caption{ Formation energies (eV/atom) of weakly stable bismuth compounds from theory and experiment.}
\begin{tabular}{|l|c|c|}
\hline 
 & E (Theory)  & E (Expt.) \tabularnewline
 & eV/atom  & eV/atom \tabularnewline
\hline 
MnBi  & -0.102  & -0.102 \tabularnewline
NiBi  & -0.064  & -0.020 \tabularnewline
Ni$_{2}$Mn$_{5}$Bi$_{4}$  & -0.117  & stable \tabularnewline
CuNi$_{3}$Mn$_{4}$Bi$_{4}$  & -0.106  & --- \tabularnewline
ZnNi$_{7}$Mn$_{8}$Bi$_{8}$  & -0.115  & --- \tabularnewline
\hline 
\end{tabular}
\label{t1E} 
\end{table}

\end{document}